\documentclass[aps,prd,preprint,onecolumn,amsfonts,amsmath,amssymb,showkeys,showpacs,a4paper,tightenlines,superscriptaddress]{revtex4-1}
\usepackage{graphicx}
\usepackage{subfig}
\usepackage{bm}
\usepackage{ulem}
\usepackage{dcolumn}
\usepackage[colorlinks,breaklinks=true,linkcolor=blue,urlcolor=blue,citecolor=blue]{hyperref}
\usepackage{cancel}
\usepackage{color}
\usepackage{appendix}

\begin{document}

\title{Signature of the interaction between dark sectors in the reionization process}

\affiliation{Center for Gravitation and Cosmology, College of Physical Science and Technology, Yangzhou University, Yangzhou 225009, China}
\affiliation{School of Physics and Astronomy, Shanghai Jiao Tong University, Shanghai 200240, China}
\affiliation{IFSA Collaborative Innovation Center, Shanghai Jiao Tong University, Shanghai 200240, China}
\affiliation{Key Laboratory for Computational Astrophysics, National Astronomical Observatories, Chinese Academy of Sciences, Beijing 100012, China}

\author{Rui An$^{2,3}$}
\email{an\_rui@sjtu.edu.cn}
\author{Xiaodong Xu$^{1}$}
\author{Jun Zhang$^{2}$}
\email{betajzhang@sjtu.edu.cn}
\author{Bin Wang$^{1,2}$}
\email{wang\_b@sjtu.edu.cn}
\thanks{Corresponding Author}
\author{Bin Yue$^{4}$}

\begin{abstract}
We consider the reionization process in a cosmological model in which dark matter interacts with dark energy. Using a semi-analytical reionization model, we compute the evolution of the ionized fraction in terms of its spatial average and linear perturbations. We show that certain types of interactions between dark matter and dark energy can significantly affect the reionization history. We calculate the 21 cm signals in the interaction models, and compare the results with the predictions of the ${\rm \Lambda CDM}$ model.

\end{abstract}

\maketitle

\section{Introduction}

It has become a well established observational fact that our universe is undergoing an accelerated expansion  \cite{Riess1998, Perlmutter1999a}. The simplest theory to account for the accelerated expansion in the framework of General Relativity(GR) is a positive cosmological constant, an old ingredient in GR since its very beginning. Before the discovery of the accelerated expansion, researchers have already encountered the missing mass problem: there is a mismatch between the dynamics and distribution of visible matter from the scale of galaxy to the scale of the cosmic microwave background(CMB) radiation \cite{Oort1932, Zwicky1933, Smith1936, Rubin1970, Rubin1980, Sofue2001, Perlmutter1999a, Clowe2006, Nolta2009}. To explain the mass discrepancy, one usually postulates the existence of dark matter(DM). The cosmological constant($\Lambda$), cold dark matter(CDM), together with normal matter and radiation constitute the matter components in the standard cosmological model, $\Lambda$CDM model. Yet, theoretically the cosmological constant suffers from some problems such as the cosmological constant problem and the coincidence problem \cite{Weinberg1989, Weinberg2007}. Researchers have explored a variety of scenarios to describe the late time accelerated expansion beyond the cosmological constant. Great amounts of such efforts involve replacing the cosmological constant in $\Lambda$CDM model with some exotic forms of matter fields with negative pressure, so-called dark energy(DE). Meanwhile, $\Lambda$CDM is also in trouble in explaining the latest observations. It is found that there is a $3\sigma$ mismatch between the Hubble constant $H_0$ measured in the direct local measurement \cite{Riess2016} and the value inferred from the CMB experiment if $\Lambda$CDM is assumed \cite{PlanckCollaboration2016}. Besides, the results presented by the Baryon Oscillation Spectroscopic Survey (BOSS) experiment of the Sloan Digital Sky Survey (SDSS) indicates a 2.5$\sigma$ deviation from $\Lambda$CDM model in the measurements of the Hubble constant $H_0$ and angular distance at an average redshift $z=2.34$ \cite{Delubac2015}. Recently, \cite{Hildebrandt2017} examines the $\Lambda$CDM model by employing weak lensing data taken from a $450-\text{deg}^2$ observing field of the Kilo Degree Survey (KiDS) and they found that there exists a `substantial discordance' inferred from the $\Lambda$CDM model between the KiDS data and the Planck 2015 CMB data. Due to these tensions between observations of large scale structure and CMB measurements, it is then reasonable to resort to dark energy models for reconciliation between the observations.

If dark energy is responsible for the accelerated expansion, our universe's energy density is dominated by dark matter and dark energy today. The natures of both remain unknown. There is no detection of non-gravitational interaction of DM or DE with baryonic matter and radiation to date. From the field theory point of view, it is interesting to consider if dark sectors, two biggest components of our universe, interact with each other. It was found that suitable interaction between DM and DE can help to alleviate the coincidence problem. Confronting to observations, it was observed that DE and DM interaction is allowed by astronomical observations. Readers can refer to \cite{Wang2016} for a review on theoretical challenges, cosmological implications and observational signatures in interacting dark energy(IDE) models. More recently, it was illustrated  that a simple phenomenological interaction in the dark sector can be a solution to the discordance between the Hubble constant and the angular distance of BOSS at redshift $z=2.34$ \cite{Ferreira2014}. For the tensions between the weak lensing measurements and the CMB measurements, it was shown in \cite{An2018} that a desired concordance between KiDS and Planck datasets can be achieved by the interacting dark energy models.

Most discussions on the interaction between DM and DE are phenomenological. It is a question to ask whether such phenomenological setups make sense in a realistic quantum field theory. There should be no significant interactions between heavy dark matter and light dark energy at the microscopic level from the normal perturbative quantum field theory \cite{D'Amico2016}.  However, it was argued in \cite{D'Amico2016} that if dark energy and a fraction of dark matter are very light axions, they can have significant mixings which are radiatively stable and perfectly consistent with quantum field theory. The cosmological evolutions in the background and perturbations have been discussed with such kind of interaction between dark sectors with quantum origin. It was shown that the description of cosmological evolutions by the quantum field theory model of interacting dark
matter and dark energy is similar to that of the phenomenological model.

More recently, the signatures of IDE models have been disclosed in various astronomical observations \cite{Amendola2003, Wang2007a, Feng2007, Feng2008, Xia2009, Martinelli2010, Valiviita2009, Honorez2010a, He2009, He2010a, Xu2011, Xu2013, Salvatelli2013, Costa2013, Salvatelli2014, Li2014a, Bertolami2007, Abdalla2007, Abdalla2009, Murgia2016, Sola2018, Li2013, Yang2014a, Xia2013, Fabris2009, Guo2007a}. Since DE is prominent at redshift $z \lesssim 2$, it is informative to test interacting DE models with observations at low redshift based on the expansion history of our universe, such as the measurement of the Hubble constant and the luminosity/angular diameter distances' tests etc, and different observations from large scale structures. Besides, considering that the interaction extends the effect to larger redshifts and could even alter the sequence of cosmological eras, it is interesting to examine the IDE models at high redshift observations. At present, the most accurate and information-rich observational data come from the experiments measuring the temperature and polarization anisotropies of CMB. From the CMB temperature anisotropies, the IDE models have been proved to be viable descriptions of the universe, see \cite{Wang2016} and references therein and also recent result in \cite{Costa2017}. \cite{Salvatelli2014} has shown that the CMB data provides a moderate Bayesian evidence in favor of an interacting vacuum model. Apart from the CMB radiation, it is also worth exploring what signals the IDE models left at redshifts that bridge the gap between accelerated expansion phase and the CMB era. It is therefore useful to extend our study and further constrain the IDE models at high redshifts.

Observationally, probes to the dark ages and epoch of reionization(EoR), ranging from $z \simeq 1100$ to $z \simeq 6$, are much less rich and accurate than low redshift observations. We have several observational evidences from which information of EoR can be inferred. The CMB experiments constrain the electron scattering optical depth to the last scattering surface\cite{PlanckCollaboration2016a}. The redshift evolution of QSO spectra and Ly$\alpha$ emitters provide hints of reionization history at $z \gtrsim 6$ \cite{Fan2006, Carilli2010, Mortlock2011, Caruana2014, Schenker2014}. Furthermore, detection of 21cm radiation signals is expected to unveil the history of the universe during EoR. The 21cm line corresponds to the transition between the fundamental hyperfine levels of neutral hydrogen(HI) atoms. Please see the recent review \cite{Kashlinsky2018} and references therein for detailed descriptions of reionization. Several ongoing 21cm radio telescopes and interferometers, including LOFAR \cite{VanHaarlem2013, Yatawatta2013}, MWA \cite{Tingay2012, Bowman2012}, PAPER \cite{Parsons2013, Jacobs2014, Ali2015},  HERA \cite{Pober2013}, BINGO \cite{Battye2013} and the upcoming SKA-LOW \footnote{http://www.skatelescope.org}, are aiming to detect the 21cm signals from the epoch of reionization. Considering the promising experimental results in the near future, it is of great interest to carry out theoretical analysis of signatures of IDE model in the reionization era in this work.

The brightness temperature fluctuation of the redshifted 21cm signal reflects the distributions of HI, which serves as a tracer of the 3D large scale structure of the universe. In order to interprete observational data and constrain cosmological models, a model of reionization on large scales is necessary. In principle, hydrodynamic simulations offer reliable way of modeling reionization process \cite{Kohler2005, Iliev2005, Iliev2006, Zahn2006, McQuinn2006, Lee2007, Shin2007, Iliev2009, Aubert2010, Friedrich2011, Iliev2014, So2014, Norman2015}. However, simulations are limited by the scale range it can cover. It is difficult for the simulations to have large box size to adopt large scale correlations while still be able to resolve the necessary small scale structure, even with the state-of-art computer capabilities. Furthermore, we need to explore various parameters in order to constrain cosmological models, thus computationally less expensive, semi-analytic models, e.g. \cite{Miralda-Escude2000, Furlanetto2004, Mesinger2007, Kuhlen2012}, are preferred. In this work, we will employ the reionization model developed in \cite{Zhang2007a} and examine the imprint of the interaction between dark sectors in this reionization model.

The organization of our paper is the following: in the next section we will review the IDE models and list its linear perturbation equations. In Sec.\ref{sec.reionization_model} and Sec.\ref{sec.21cm_power_spectrum}, we will go over the semi-analytical reionization description and explain the way to calculate the 21cm radiation power spectrum. In Sec.\ref{sec.results}, we will report how the reionization process and the 21cm radiation power spectrum will be modified if there is interaction between DE and DM. Our conclusion and discussion will be given in the final section.

\section{Interacting dark energy model \label{sec.IDE_model}}

In the interacting DM-DE scenario, the energy momentum tensor of DM and DE are not conserved separately, but
\begin{equation}
\nabla_{\mu} T_{(\lambda)}^{\mu\nu} = Q_{(\lambda)}^{\nu},
\label{eq.EM_conservation}
\end{equation}
where the subscript `$\lambda$' refers to either DM(`$\mathrm{c}$') or DE(`$\mathrm{d}$'). $Q_{(\lambda)}^{\nu}$ is the coupling vector representing the interaction between DM and DE. We assume that the dark sectors do not interact with normal matter in addition to gravitational interactions, thus the total energy momentum of dark sectors is conserved, so that $Q_{\mathrm{c}} + Q_{\mathrm{d}} = 0$.

The universe is described by a flat Friedmann-Lemaitre-Robertson-Walker(FLRW) metric with small perturbations over a smooth background. The line element is given by
\begin{equation}
\mathrm{d}s^2 = a^2 [ -(1+2\psi)\mathrm{d}\tau^2 + 2\partial_iB\mathrm{d}\tau\mathrm{d}x^i + (1+2\phi)\delta_{ij}\mathrm{d}x^i\mathrm{d}x^j + (\partial_i\partial_j - \frac{1}{3}\delta_{ij}\nabla^2)E\mathrm{d}x^i\mathrm{d}x^j ],
\end{equation}
where we have only considered the scalar perturbations to the metric ($\psi$, $B$, $\phi$ and $E$). In the following, we work in Newtonian gauge such that $B=E=0$.

The matter components in the universe are described by the energy momentum tensor of ideal fluid
\begin{equation}
T^{\mu\nu} = (\rho+p) U^{\mu}U^{\nu} + pg^{\mu\nu}.
\end{equation}
Given \eqref{eq.EM_conservation}, we obtain the equations of motions for the DM/DE energy densities in the homogeneous and isotropic background
\begin{align}
& \dot{\rho_c} + 3\frac{\dot{a}}{a}\rho_c = a^2Q_{\mathrm{c}}^0 = aQ_{\mathrm{c}}, \label{eq.rhoc}\\
& \dot{\rho_d} + 3\frac{\dot{a}}{a}(1+w)\rho_d = a^2Q_{\mathrm{d}}^0 = aQ_{\mathrm{d}}, \label{eq.rhod}
\end{align}
A dot denotes the derivative with respect to conformal time. $w$ is the equation of state of DE. Due to lack of understandings about either dark matter or dark energy at present, it is difficult to postulate the coupling between them from first principles. Hence we would instead describe the interaction phenomenologically and focus on its impact on the dynamics of the universe rather than the microscopic mechanism. The interaction between DM and DE represents only a small correction to the evolution of the Universe. In particle physics, we expect the interaction to be a function of the energy densities, $\rho_c, \rho_d$ and the inverse of the Hubble constant, $H^{-1}$. To the first order of Taylor expansion, we obtain the form of the interaction kernel
\begin{equation}
Q_{\mathrm{c}} = -Q_{\mathrm{d}} = 3H(\xi_1\rho_{\mathrm{c}} + \xi_2\rho_{\mathrm{d}}).
\label{eq.Q}
\end{equation}
$\xi_1$ and $\xi_2$ are the dimensionless coupling coefficients, which we assume to be constants. While it is a simple parametrization of DM-DE interaction, we can put some constraints on the coupling constants as well as DE EoS $w$ by some physical considerations. First, to avoid the unphysical solution of a negative dark energy density ($\rho_{\mathrm{d}} < 0$) in the early universe, the coupling constant $\xi_1$ must be positive \cite{He2008}. Furthermore, it has been found that, when $\xi_1 \neq 0$ and $w$ is a constant greater than $-1$, the curvature perturbation diverges in early times \cite{He2008a, Xu2011, Gavela2009}. Thus we exclude such cases in our study. The linear perturbation to the zeroth component of the coupling vector can be derived from \eqref{eq.Q} \cite{He2010a}:
\begin{equation}
\delta Q_{(\lambda)}^0 = -\frac{\psi}{a}Q_{(\lambda)} + \frac{1}{a}\delta Q_{(\lambda)},
\end{equation}
while the $i$th component needs to be specified in the model in addition to the background energy transfer. In this work, we assume $\delta Q_{(\lambda)}^i$ vanishes \cite{He2010a}, i.e., there is no scattering between DM and DE, and only an inertial drag effect appears due to stationary energy transfer.

In the Newtonian gauge, the perturbed energy momentum nonconservation equation \eqref{eq.EM_conservation} leads to, on the first order, \cite{He2010a}
\begin{align}
\dot{\delta_{\mathrm{c}}} =& -kv_{\mathrm{c}} - 3\dot{\phi} + 3\frac{\dot{a}}{a}(\xi_1 + \xi_2/r_{\mathrm{D}})\psi + 3\frac{\dot{a}}{a}\xi_2(\delta_{\mathrm{d}}-\delta_{\mathrm{c}})/r_{\mathrm{D}}, \label{eq.deltac2}\\
\dot{v_{\mathrm{c}}} =& -\frac{\dot{a}}{a}v_{\mathrm{c}} + k\psi - 3\frac{\dot{a}}{a}(\xi_1 + \xi_2/r_{\mathrm{D}})v_{\mathrm{c}}, \label{eq.vc2}\\
\dot{\delta_{\mathrm{d}}} =& -3\frac{\dot{a}}{a}(c_{\mathrm{e}}^2 - w)\delta_{\mathrm{d}} - 9\big(\frac{\dot{a}}{a}\big)^2(c_{\mathrm{e}}^2 - c_{\mathrm{a}}^2)(1 + w)\frac{v_{\mathrm{d}}}{k} - (1+w)kv_{\mathrm{d}} - 3(1+w)\dot{\phi} \nonumber \\
  &- 9\big(\frac{\dot{a}}{a}\big)^2(c_{\mathrm{e}}^2 - c_{\mathrm{1}}^2)(\xi_1r_{\mathrm{D}} + \xi_2)\frac{v_{\mathrm{d}}}{k} - 3\frac{\dot{a}}{a}(\xi_1r_{\mathrm{D}} + \xi_2)\psi + 3\frac{\dot{a}}{a}\xi_1r_{\mathrm{D}}(\delta_{\mathrm{d}} -\delta_{\mathrm{c}}), \label{eq.deltad2}\\
\dot{v_{\mathrm{d}}} =& -\frac{\dot{a}}{a}(1 - 3w)v_{\mathrm{d}} + \frac{k}{1+w}c_{\mathrm{e}}^2\delta_{\mathrm{d}} + 3\frac{\dot{a}}{a}(c_{\mathrm{d}}^2 - c_{\mathrm{a}}^2)v_{\mathrm{d}} - \frac{\dot{w}}{1+w}v_{\mathrm{d}} + k\psi \nonumber \\
  &+ 3\frac{\dot{a}}{a}(c_{\mathrm{e}}^2 - c_{\mathrm{a}}^2)(\xi_1r_{\mathrm{D}} + \xi_2)\frac{v_{\mathrm{d}}}{1+w} + 3\frac{\dot{a}}{a}(\xi_1r_{\mathrm{D}} + \xi_2)v_{\mathrm{d}}, \label{eq.vd2}
\end{align}
where $r_{\mathrm{D}} \equiv \rho_{\mathrm{c}}/\rho_{\mathrm{d}}$; $c_{\mathrm{a}} \equiv \frac{\dot{p}_{\mathrm{d}}}{\dot{\rho_{\mathrm{d}}}}$ is the adiabatic sound speed and $c_{\mathrm{e}} \equiv \frac{\delta p_{\mathrm{d}}}{\delta \rho_{\mathrm{d}}}$ is the effective sound speed of DE in its rest frame.

As DM and DE do not interact with normal matter except through their gravitational effect, the physics of baryonic matter remains the same as in $\Lambda$CDM model. In particular, the equations of motion for its mean energy density and linear perturbations read
\begin{equation}
\dot{\rho}_{\mathrm{b}} + 3\frac{\dot{a}}{a}\rho_{\mathrm{b}} = 0,
\label{eq.rhob}
\end{equation}
and
\begin{align}
\dot{\delta}_{\mathrm{b}} =& -kv_{\mathrm{b}} - 3\dot{\phi}, \label{eq.deltab}\\
\dot{v}_{\mathrm{b}} =& -\frac{\dot{a}}{a}v_{\mathrm{b}} + k\psi. \label{eq.vb}
\end{align}

\section{Reionization model \label{sec.reionization_model}}

In this section, we briefly review the semi-analytical model developed in \cite{Zhang2007a}, which predicts the power spectrum of the HII fluctuations on large scales by solving the equations of ionization balance and radiative transfer to the first order in perturbations.

First, the equation of ionization equilibrium reads
\begin{equation}
\frac{\partial{n_{\mathrm{HII}}}}{\partial \tau} + \nabla \cdot (n_{\mathrm{HII}}\mathbf{u}) = (n_{\mathrm{H}} - n_{\mathrm{HII}}) \int^{\infty}_{0} \mathrm{d}\mu \int \mathrm{d}\Omega^2 n_{\gamma} \frac{\sigma(\mu)}{a^2(\tau)} \kappa(\mu,\phi) - \frac{\alpha_B n_{\mathrm{HII}}^2}{a^2(\tau)},
\label{eq.ionization}
\end{equation}
where $\tau$ is the conformal time; $n_{\mathrm{H}}$ and $n_{\mathrm{HII}}$ are the comoving number densities of the total and ionized hydrogen atoms; $n_{\gamma} = n_{\gamma}(\mathbf{x}, \tau, \mu, \mathbf{\Omega})$ is the comoving photon number density per unit volume, per unit conformal time, per unit frequency parameter and per unit solid angle. $\mathbf{\Omega}$ is the unit vector along the direction of photon propagation. The frequency parameter is defined as $\mu = \ln \nu - \ln \nu_0$, where $\nu$ is the photon's frequency and $\nu_0 = 13.6\mathrm{eV} / (2\pi \hbar)$. $\sigma(\mu)$ is the photon ionization cross section \cite{Osterbrock2006}. $\alpha_B = 2.6 \times 10^{-13} \mathrm{cm}^3\mathrm{s}^{-1}$ is the case B recombination coefficient at temperature equaling to $10^4$K. And $a$ is the scale factor. The factor $\kappa(\mu, \phi) = 1 + C(\exp(\mu) - 1) (1 - \phi^a)^b$ accounts for multiple ionizations by X-ray photons through secondary ionizations by the fast photoelectrons, where $\phi = n_{\mathrm{HII}}/n_{\mathrm{H}}$ is the local ionization fraction, and we adopt the parameters: $C=0.3908$, $a=0.4092$ and $b=1.7592$, following \cite{Shull1985}. The evolution of the radiation background is described by the radiative transfer equation
\begin{equation}
\frac{\partial n_{\gamma}}{\partial \tau} + \mathbf{\Omega} \cdot \nabla n_{\gamma} - H(\tau)a(\tau)\frac{\partial n_{\gamma}}{\partial \mu} = \frac{S}{4\pi} - (n_{\mathrm{H}} - n_{\mathrm{HII}}) n_{\gamma} \frac{\sigma(\mu)}{a^2(\tau)},
\label{eq.radiative_transfer}
\end{equation}
$S(\mathbf{x}, \tau, \mu, \mathbf{\Omega})/4\pi$ is the differential ionizing emissivity which gives the number of photons emitted by sources per unit volume, per unit conformal time, per unit frequency parameter and per unit solid angle. \eqref{eq.radiative_transfer} includes the effects of sources, photon ionization process, diffusion of photons and redshift due to the expansion of the universe. Throughout the paper, we ignore the ionization of helium atoms and assume electric neutrality everywhere in the universe.

To solve the ionization equilibrium and radiative transfer equations, we need to specify the emissivity function and the distribution and spectrum of ionizing sources. In this work, we employ the extended Press-Schechter formalism \cite{Press1974, Bond1991, Lacey1993} to model the ionizing sources. The minimal mass of a halo which can host luminous sources is given by
\begin{equation}
M_{\mathrm{min}} \approx 1.3 \times 10^7 M_{\odot} \bigg(\frac{T_{\mathrm{vir}}}{10^4 K}\bigg)^{3/2} \bigg(\frac{1+z}{21}\bigg)^{-3/2} \bigg(\frac{\Omega_{m}}{0.3}\bigg)^{-1/2} \bigg(\frac{h}{0.7}\bigg)^{-1} \bigg(\frac{\mu_{\mathrm{mol}}}{1.22}\bigg)^{-3/2},
\end{equation}
where $\mu_{\mathrm{mol}}$ is the mean molecular weight. On mass scale $m$, the fraction of mass collapsed in halos with masses larger than $M_{\mathrm{min}}$ is
\begin{equation}
f^{\mathrm{coll}}_{\mathrm{m}}(\mathbf{x}, \tau) = \mathrm{erfc}\bigg[ \frac{\delta_{\mathrm{cri}} - \delta_{\mathrm{m}}(\mathbf{x}, \tau)}{\sqrt{2[\sigma_{\mathrm{min}}^2(\tau) - \sigma_{\mathrm{m}}^2(\tau)]}} \bigg],
\end{equation}
where $\delta_{\mathrm{cri}}$ is the critical overdensity in the spherical collapse model; $\delta_{\mathrm{m}}$ and $\sigma_{\mathrm{m}}$ are the overdensity and variance of the density fluctuations on mass scale $m$ and $\sigma_{\mathrm{min}}$ is the variance of density fluctuation corresponding to mass scale $M_{\mathrm{min}}$. In the following, we neglect $\sigma_{\mathrm{m}}$ since it is much smaller than $\sigma_{\mathrm{min}}$ on large scales. Smoothed over mass scale $m$, the emissivity function reads
\begin{equation}
S_{\mathrm{m}}(\mathbf{x}, \mu, \tau) = \gamma(\mu) \bar{n}_{\mathrm{H}} \frac{\partial}{\partial \tau}[f^{\mathrm{coll}}_{\mathrm{m}}(\mathbf{x}, \tau) (1 + \delta_{\mathrm{b}}(\mathbf{x}, \tau))].
\label{eq.emissivity}
\end{equation}
$\gamma(\mu)$ represents the average number of ionizing photons emitted by each hydrogen atom in the collapsed objects per unit frequency parameter $\mu$.

We can rewrite the number densities and emissivity in terms of their spatially averages and perturbations:
\begin{align}
n_{\mathrm{HII}} &= \bar{n}_{\mathrm{H}} f_{\mathrm{HII}}(\tau) [1+\delta_{\mathrm{HII}}(\mathbf{x}, \tau)] = \bar{n}_{\mathrm{H}}[ f_{\mathrm{HII}}(\tau) + \Delta_{\mathrm{HII}}(\mathbf{x}, \tau)], \\
n_{\mathrm{H}} &= \bar{n}_{\mathrm{H}}[1+\delta_{\mathrm{b}}(\mathbf{x}, \tau)], \\
n_{\gamma} &= \bar{n}_{\mathrm{H}} f_{\gamma}(\tau, \mu) [1+\delta_{\gamma}(\mathbf{x}, \tau, \mu, \mathbf{\Omega})] = \bar{n}_{\mathrm{H}}[ f_{\gamma}(\tau, \mu) + \Delta_{\gamma}(\mathbf{x}, \tau, \mu, \mathbf{\Omega})] \\
S &= \bar{n}_{\mathrm{H}} f_{\mathrm{s}}(\tau, \mu) [1+\delta_{\mathrm{s}}(\mathbf{x}, \tau, \mu, \mathbf{\Omega})] = \bar{n}_{\mathrm{H}}[ f_{\mathrm{s}}(\tau, \mu) + \Delta_{\mathrm{s}}(\mathbf{x}, \tau, \mu, \mathbf{\Omega}],
\end{align}
where $f_{\mathrm{HII}}$ and $f_{\gamma}$ are the ratio of mean ionized hydrogen atoms and photons number densities to the average total hydrogen number density $\bar{n}_{\mathrm{H}}$, and $f_{\mathrm{s}}$ is the mean source emissivity normalized by $\bar{n}_{\mathrm{H}}$. We assume that the overdensity of the total hydrogen atoms is the same as the baryon overdensity on large scales. From \eqref{eq.ionization} and \eqref{eq.radiative_transfer}, we find the equations of the spatially averaged quantities
\begin{align}
\frac{\partial f_{\mathrm{HII}}}{\partial \tau} &= 4\pi(1 - f_{\mathrm{HII}}) \int \mathrm{d}\mu \frac{\sigma \bar{n}_{\mathrm{H}}}{a^2} \langle\kappa\rangle f_{\gamma} C_{\gamma\mathrm{H}}^{(1)} - \frac{\alpha_{\mathrm{B}}\bar{n}_{\mathrm{H}}}{a^2} f_{\mathrm{HII}}^2 C_{\mathrm{HII}}, \label{eq.ionization_mean} \\
\frac{\partial f_{\gamma}}{\partial \tau} &= \frac{f_{\mathrm{s}}}{4\pi} + Ha\frac{\partial f_{\gamma}}{\partial\mu} - \frac{\sigma \bar{n}_{\mathrm{H}}}{a^2} (1 - f_{\mathrm{HII}}) f_{\gamma} C_{\gamma\mathrm{H}}^{(2)}, \label{eq.radiative_transfer_mean}
\end{align}
where $C_{\gamma\mathrm{H}}^{(1)}$, $C_{\gamma\mathrm{H}}^{(2)}$ and $C_{\mathrm{HII}}$ are the clumping factors for photoionization and recombination, respectively:
\begin{equation}
C_{\gamma\mathrm{H}}^{(1)} \equiv \frac{\langle n_{\mathrm{HI}} n_{\gamma} \kappa \rangle}{\langle n_{\mathrm{HI}} \rangle \langle n_{\gamma} \rangle \langle \kappa \rangle}, ~ C_{\gamma\mathrm{H}}^{(2)} \equiv \frac{\langle n_{\mathrm{HI}} n_{\gamma} \rangle}{\langle n_{\mathrm{HI}} \rangle \langle n_{\gamma} \rangle}, ~ C_{\mathrm{HII}} \equiv \frac{\langle n_{\mathrm{HII}}^2 \rangle}{\langle n_{\mathrm{HII}} \rangle^2}.
\end{equation}

In Fourier space, the linear perturbations of \eqref{eq.ionization} and \eqref{eq.radiative_transfer} lead to
\begin{align}
\frac{\partial \tilde{\Delta}_{\mathrm{HII}}}{\partial \omega} &= G\tilde{\delta_{\mathrm{b}}} - F\tilde{\Delta}_{\mathrm{HII}} + \int_0^{\infty} \mathrm{d}\mu \langle\kappa\rangle \int \mathrm{d}^2\mathbf{\Omega} \tilde{\Delta}_{\gamma}B, \label{eq.ionization_pert} \\
\frac{\partial \tilde{\Delta}_{\gamma}}{\partial \omega} &= \frac{\partial \tilde{\Delta}_{\gamma}}{\partial \mu} - M\tilde{\Delta}_{\gamma} + N\tilde{\Delta}_{\mathrm{s}} + R(\tilde{\Delta}_{\mathrm{HII}} - \tilde{\delta}_{\mathrm{b}}), \label{eq.radiative_transfer_pert}
\end{align}
where $\tilde{\Delta}_{\mathrm{HII}}$, $\tilde{\Delta}_{\gamma}$, $\tilde{\Delta}_{\mathrm{s}}$ and $\tilde{\delta}_{\mathrm{b}}$ are the Fourier transforms of $\Delta_{\mathrm{HII}}$, $\Delta_{\gamma}$, $\Delta_{\mathrm{s}}$ and $\delta_{\mathrm{b}}$, and
\begin{align*}
F &= 2\tilde{\alpha}_{\mathrm{B}}f_{\mathrm{HII}} + 4\pi \int_{0}^{\infty}\mathrm{d}\mu \tilde{\sigma}f_{\gamma} \bigg[ \langle\kappa\rangle - (1-f_{\mathrm{HII}}) \frac{\partial\kappa}{\partial\phi}\bigg\vert_{\phi=f_{\mathrm{HII}}} \bigg], \\
G &= \frac{\mathrm{d}\ln D_{\mathrm{b}}}{\mathrm{d}\omega} f_{\mathrm{HII}} + 4\pi \int_{0}^{\infty}\mathrm{d}\mu \tilde{\sigma}f_{\gamma} \bigg[ \langle\kappa\rangle - (1-f_{\mathrm{HII}})f_{\mathrm{HII}} \frac{\partial\kappa}{\partial\phi}\bigg\vert_{\phi=f_{\mathrm{HII}}} \bigg], \\
B &= (1 - f_{\mathrm{HII}}) \tilde{\sigma}, \\
M &= (1 - f_{\mathrm{HII}}) \tilde{\sigma} - \frac{i\mathbf{k}\cdot\mathbf{\Omega}}{Ha}, \\
N &= (4\pi Ha)^{-1}, \\
R &= \tilde{\sigma}f_{\gamma},
\end{align*}
where $D_{\mathrm{b}}$ is the linear growth factor of baryon. For simplicity, we define $\tilde{\alpha}_{\mathrm{B}}(\tau) \equiv \alpha_{\mathrm{B}}\bar{n}_{\mathrm{H}} / (Ha^3)$ and $\tilde{\sigma}(\mu, \tau) \equiv \sigma\bar{n}_{\mathrm{H}} / (Ha^3)$.

Taking the large $m$ limit of \eqref{eq.emissivity}, we find the mean emissivity function
\begin{equation}
f_{\mathrm{s}}(\mu, \tau) = \gamma(\mu) \frac{2}{\sqrt{\pi}} \exp\bigg( -\frac{\delta_{\mathrm{c}}^2}{2\sigma_{\mathrm{min}}^2} \bigg) \frac{\mathrm{d}}{\mathrm{d}\tau}\bigg( -\frac{\delta_{\mathrm{c}}}{\sqrt{2\sigma_{\mathrm{min}}^2}} \bigg).
\end{equation}
The linear perturbation to the emissivity is obtained by doing Tayler expansion of \eqref{eq.emissivity} around $\delta_{\mathrm{b}}$:
\begin{equation}
\tilde{\Delta}_{\mathrm{s}}(\mathbf{k}, \mu, \tau) = \gamma(\mu) \frac{\partial}{\partial\tau} [ R(\tau)\tilde{\delta}_{\mathrm{b}}(\mathbf{k}, \tau) ],
\end{equation}
where
\begin{equation}
R(\tau) = \mathrm{erfc}\bigg(\frac{\delta_{\mathrm{c}}}{\sqrt{2\sigma_{\mathrm{min}}^2}}\bigg) + \sqrt{\frac{2}{\pi\sigma_{\mathrm{min}}^2}} \exp\bigg(-\frac{\delta_{\mathrm{c}}^2}{2\sigma_{\mathrm{min}}^2}\bigg).
\end{equation}

Numerically solving \eqref{eq.ionization_mean}, \eqref{eq.radiative_transfer_mean}, \eqref{eq.ionization_pert} and \eqref{eq.radiative_transfer_pert}, we can calculate the linear power spectrum of ionized hydrogen atoms.

\section{21cm radiation power spectrum \label{sec.21cm_power_spectrum}}

During the reionization epoch, the absorption and re-emission of Lyman-$\alpha$ photons and nonlinear effect determine the spin temperature of hydrogen atoms, which defines the relative abundance of triplet and singlet hydrogen states. When the spin temperature departs from the CMB temperature, a net absorption or emission of the 21 cm line does appear against the CMB radiation. The difference between the observed 21 cm brightness temperature and the CMB temperature is given by\cite{Lewis2007}
\begin{equation}
T_\mathrm{b} = \frac{3hc^3 A_{10} n_{\mathrm{HI}}}{32\pi k_{\mathrm{B}} \nu_0^2 (1+z)^2 (\mathrm{d}u_{\Vert} / dr)} \frac{T_{\mathrm{s}} - T_{\mathrm{CMB}}}{T_{\mathrm{s}}},
\end{equation}
where $A_{10}$ is the spontaneous decay rate of 21 cm transition, $n_{\mathrm{HI}}$ is the number density of the neutral hydrogen, and $\nu_0$ is the frequency of the 21 cm transition at the rest frame. $T_{\mathrm{s}}$ is the spin temperature. The relation between $T_{\mathrm{s}}$ and the populations of the hyperfine levels is
\begin{equation}
\frac{n_1}{n_0} = \frac{g_1}{g_0} \mathrm{e}^{-h\nu_0 / k_{\mathrm{B}} T_{\mathrm{s}}}.
\end{equation}
$g_0=1$ and $g_1=3$ are the statistical weights. The 21 cm signal can be observed in the form of an absorption line or an emission line against the CMB blackbody spectrum, depending on whether the spin temperature is lower or higher than the CMB temperature. $\mathrm{d}u_{\Vert} / dr$ is the gradient of the physical velocity along the line of sight and $r$ is the comoving distance.

The gradient of physical velocity can be split into
\begin{equation}
\frac{\mathrm{d}u_{\Vert}}{\mathrm{d}r} = \frac{H(z)}{1+z} + \frac{\partial v}{\partial r},
\end{equation}
where $v$ is the peculiar velocity, $H(z)$ is the Hubble parameter at given redshift. On the background level, the peculiar velocity vanishes and $\mathrm{d}u_{\Vert} / \mathrm{d}r = H(z) / (1+z)$.

The average temperature difference is
\begin{equation}
\bar{T}_{\mathrm{b}} = \frac{3hc^3 A_{10} \bar{n}_{\mathrm{HI}}}{32\pi k_{\mathrm{B}} \nu_0^2} \frac{1}{(1+z)H(z)} \frac{\bar{T}_{\mathrm{s}} - T_{\mathrm{CMB}}}{\bar{T}_{\mathrm{s}}}.
\end{equation}
During the reionization era, we assume $T_{\mathrm{s}} \gg T_{\mathrm{CMB}}$, thus $(T_{\mathrm{s}}-T_{\mathrm{CMB}})/T_{\mathrm{s}} \simeq 1$. The number density of neutral hydrogen can be written as $\bar{n}_{\mathrm{HI}} = f_{\mathrm{HI}} \bar{n}_{\mathrm{b}} (1-f_{\mathrm{He}})$, where $\bar{n}_{\mathrm{b}}$ is the average baryon number density, $f_{\mathrm{HI}}$ is the average neutral fraction and $f_{\mathrm{He}}$ is the helium fraction.

Perturbing the temperature difference, we get
\begin{equation}
T_{\mathrm{b}}(\mathbf{x}) =  \frac{\bar{T}_{\mathrm{b}}}{f_{\mathrm{HI}}} [1-f_{\mathrm{HII}}(1+\delta_{\mathrm{HII}})] (1+\delta_{\mathrm{b}}) (1-\delta_v).
\label{Tb_fluc}
\end{equation}
The gradient of the peculiar velocity is defined as $\delta_v \equiv \frac{\partial v}{\partial r} / \frac{H(z)}{1+z}$. In linear perturbation theory, we can write
\begin{equation}
\tilde{\delta}_v(\mathbf{k}) = - \frac{\mathrm{d}\ln D_{\mathrm{b}}}{\mathrm{d}\ln a} (\hat{\mathbf{n}} \cdot \hat{\mathbf{k}})^2 \tilde{\delta}_{\mathrm{b}} \equiv -\tilde{\mu}^2 \tilde{\delta}_{\mathrm{b}},
\label{delta_v}
\end{equation}
where $\hat{\mathbf{n}}$ is the unit vector along the line of sight and $\tilde{\mu}^2 \equiv \frac{\mathrm{d}\ln D_{\mathrm{b}}}{\mathrm{d}\ln a} (\hat{\mathbf{n}} \cdot \hat{\mathbf{k}})^2$. $D_{\mathrm{b}}$ is the baryon growth function.

It is now straightforward to derive the 21 cm radiation power spectrum, which is defined by
\begin{equation}
\langle \tilde{T}_{\mathrm{b}}^{\ast}(\mathbf{k}) \tilde{T}_{\mathrm{b}}(\mathbf{k}') \rangle \equiv (2\pi)^3 \delta^3(\mathbf{k}-\mathbf{k}') P_{\Delta T}(\mathbf{k}).
\end{equation}
$\tilde{T}_{\mathrm{b}}$ is the Fourier mode of the fluctuation of the brightness temperature. Using (\ref{delta_v}) and (\ref{Tb_fluc}), one can then compute the power spectrum as \cite{Mao2008}:
\begin{equation}
P_{\Delta T}(\mathbf{k}) = \bar{T}_{\mathrm{b}}^2 \{ (P_{\mathrm{bb}} - 2x P_{\mathrm{ib}} + x^2 P_{\mathrm{ii}}) + 2\tilde{\mu}^2(P_{\mathrm{bb}} - x P_{\mathrm{ib}}) + \tilde{\mu}^4 P_{\mathrm{bb}} \},
\end{equation}
where $x=f_{\mathrm{HII}}/f_{\mathrm{HI}}$, and $P_{\mathrm{bb}}$, $P_{\mathrm{ii}}$ and $P_{\mathrm{ib}}$ are the power spectrum of baryon density fluctuation, ionization fraction fluctuation and density-ionization cross correlation. $P_{\Delta T}$ is composed of three components, which have distinct angular dependence. We refer to them as
\begin{eqnarray}
&& P_0 = \bar{T}_{\mathrm{b}}^2 ( P_{\mathrm{bb}} - 2x P_{\mathrm{ib}} + x^2 P_{\mathrm{ii}}), \\
&& P_2 = \bar{T}_{\mathrm{b}}^2 2\tilde{\mu}^2(P_{\mathrm{bb}} - x P_{\mathrm{ib}}), \\
&& P_4 = \bar{T}_{\mathrm{b}}^2 \tilde{\mu}^4 P_{\mathrm{bb}}.
\label{eq.P_4}
\end{eqnarray}

\section{Results \label{sec.results}}

\subsection{Model analysis}

Using the method described in Sec.\ref{sec.reionization_model}, we calculate the reionization history in the presence of DM-DE interaction. For a comparison, we repeat the calculation for the fiducial $\Lambda$CDM model. The cosmological parameters for the fiducial $\Lambda$CDM model as well as the interacting DM-DE models are: today's Hubble constant $H_0 = 67 \mathrm{km}\cdot\mathrm{s}^{-1}\cdot\mathrm{Mpc}^{-1}$, baryon and dark matter abundance $\Omega_{\mathrm{b}}h^2 = 0.022$, $\Omega_{\mathrm{c}}h^2 = 0.12$, and the optical depth to the begining of reionization $\tau_{\mathrm{T}} = 0.08$. Besides, we assume the expansion of the universe and the evolution of fluctuations of different energy components in interacting DM-DE models are close to those of the fiducial $\Lambda$CDM model in sufficiently early time, so that we set the Hubble parameter and linear perturbations of DM and baryons the same as $\Lambda$CDM model at the decoupling $a \sim 10^{-3}$.

In order to examine the influence of DE, we will first present the result for non-interacting DE models (NIDE-W) with different DE equation of state. To be specific, we will consider two cases here: the constant equation of state $w=w_0$ and the time-dependent one which is discribed by a particular parametrization $w(a)=w_0+w_a(1-a)$. Then we will extend our discussions to interacting DE models: with the interaction between dark sectors either in proportional to energy density of DM (IDE-A) or DE (IDE-B). In the computation, we set clumping factors
\begin{equation}
C_{\gamma\mathrm{H}}^{(1)} = C_{\gamma\mathrm{H}}^{(2)} = 1, ~ C_{\mathrm{HII}} = 10,
\end{equation}
and the source spectrum
\begin{equation}
\gamma(\mu)\mathrm{d}\mu = \frac{\zeta}{C_{\beta}} \mathrm{e}^{\beta+1} \mathrm{d}\mu,
\end{equation}
where $\beta$ is the spectral index, $\zeta$ represents the total number of ionizing photons generated per baryon in stars, which manage to escape into the IGM, and $C_{\beta}$ is a normalization factor. Thus the source spectrum takes a power law form in frequency. For simplicity, we assume $\beta = -3$ and tune $\zeta$ for each model in order to satisfy $\tau_{\mathrm{T}}=0.08$. The model parameters we use in our computations are listed in Table.\ref{tab.parameters_w} and Table.\ref{tab.parameters_ide} respectively.

\begin{table*}[ht]
\caption{\label{tab.parameters_w} Parameters of the non-interacting models.}
\begin{tabular}{p{65pt}p{40pt}p{40pt}p{40pt}}
    \toprule
     & $w_0$ & $w_a$ & $\zeta$ \\
    \hline
    $\Lambda$CDM & -1.0 & - & 84.2 \\
    NIDE-W1 & -0.9 & - & 84.3 \\
    NIDE-W2 & -1.1 & - & 84.1 \\
    NIDE-W3 & -0.9 & 0.3 & 85.9 \\
    NIDE-W4 & -1.1 & -0.3 & 84.0  \\
    \lasthline
\end{tabular}
\end{table*}

\begin{table*}[ht]
\caption{\label{tab.parameters_ide} Parameters of the interacting models with constant $w$.}
\begin{tabular}{p{50pt}p{40pt}p{40pt}p{40pt}p{40pt}}
    \toprule
     & $w$ & $\xi_1$ & $\xi_2$ & $\zeta$ \\
    \hline
    $\Lambda$CDM & -1 & - & - & 84.2 \\
    IDE-A1 & -1.1 & 0.005 & 0     & 156.5 \\
    IDE-A2 & -1.1 & 0.003& 0     & 115.6 \\
    IDE-B1 & -0.9 & 0    & 0.05  & 84.8  \\
    IDE-B2 & -0.9 & 0    & -0.3  & 83.8  \\
    IDE-B3 & -1.1 & 0    & 0.3   & 87.2  \\
    IDE-B4 & -1.1 & 0    & -0.05 & 84.0  \\
    \lasthline
\end{tabular}
\end{table*}

\begin{figure*}
\subfloat[]{
    \includegraphics[width=0.45\textwidth]{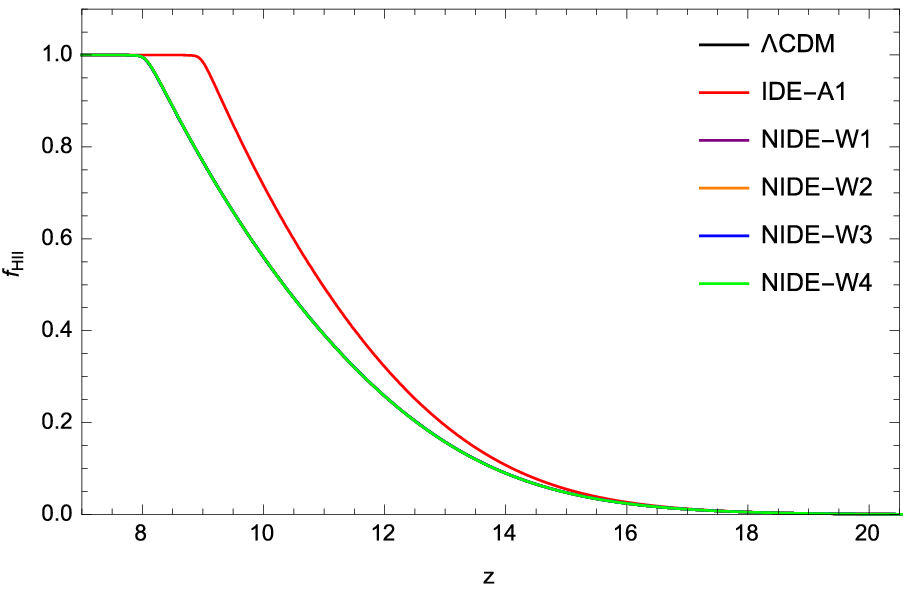}
}
\subfloat[]{
    \includegraphics[width=0.45\textwidth]{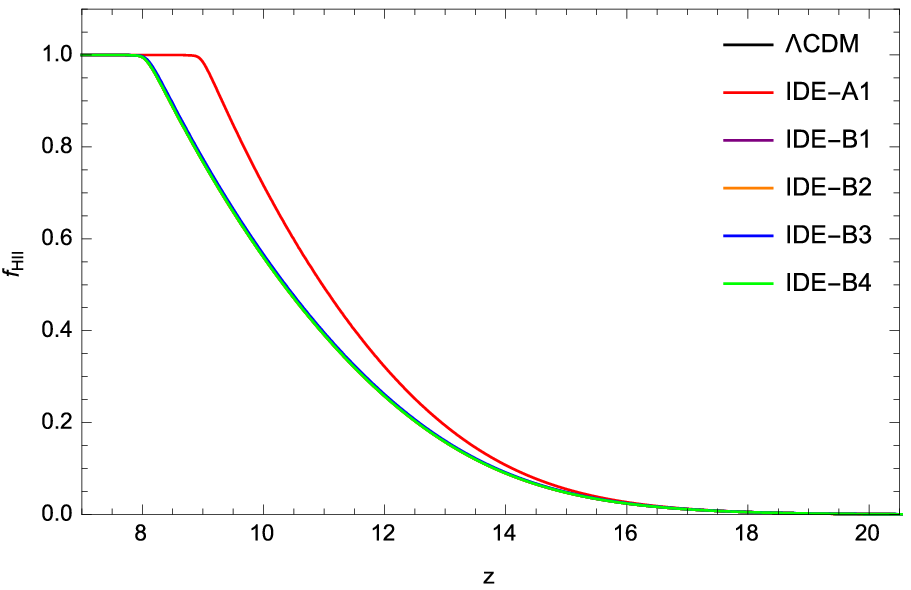}
}
\caption{\label{fig.f_HII} The evolution of ionized fraction $f_{\mathrm{HII}}$.}
\end{figure*}

\begin{figure*}
\subfloat[]{
    \includegraphics[width=0.45\textwidth]{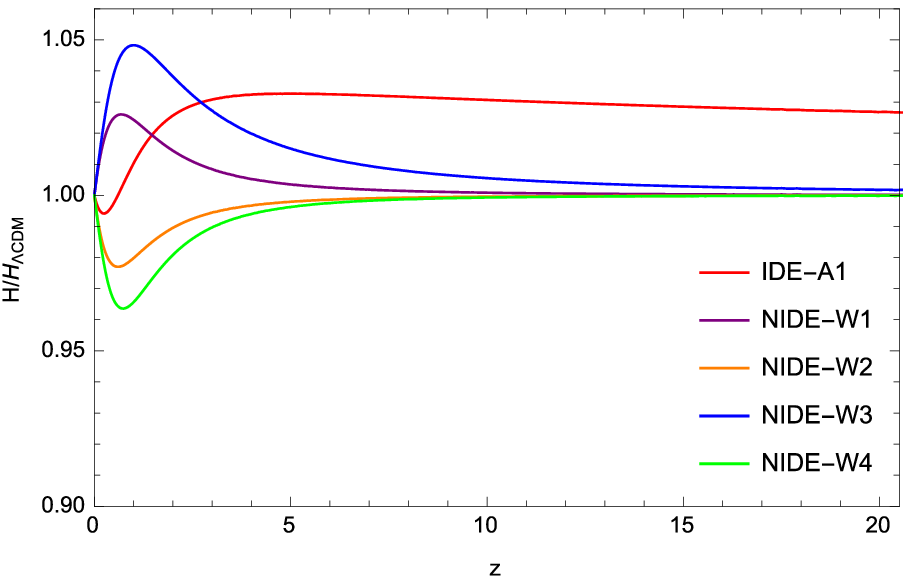}
}
\subfloat[]{
    \includegraphics[width=0.45\textwidth]{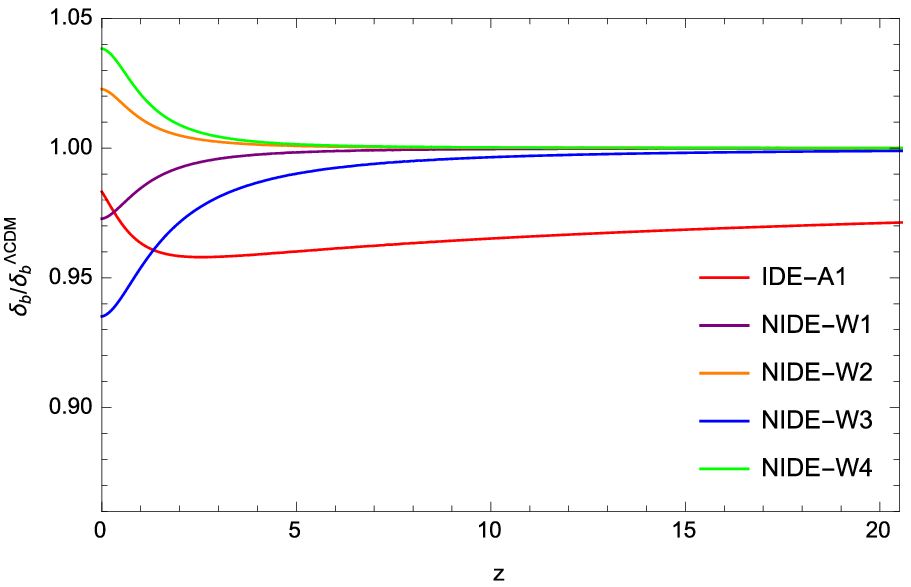}
} \\
\subfloat[]{
    \includegraphics[width=0.45\textwidth]{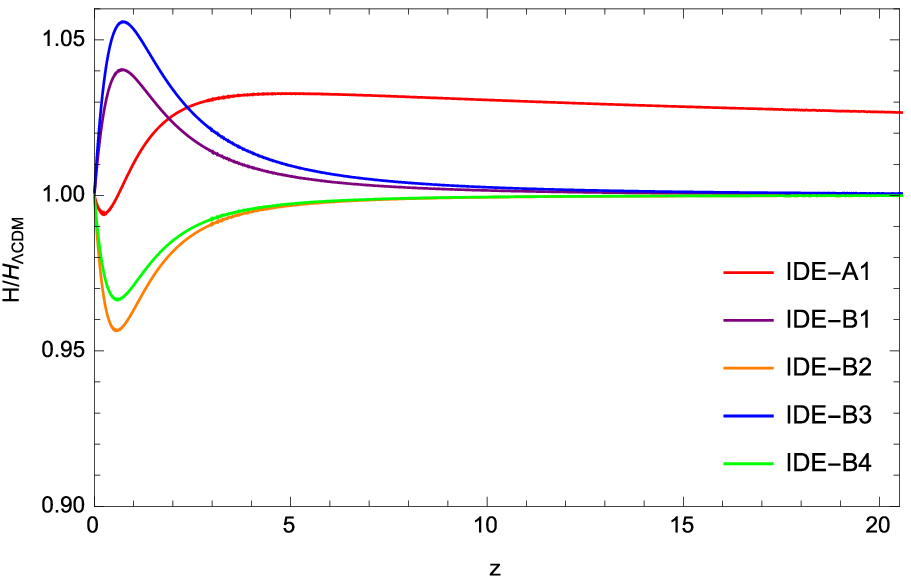}
}
\subfloat[]{
    \includegraphics[width=0.45\textwidth]{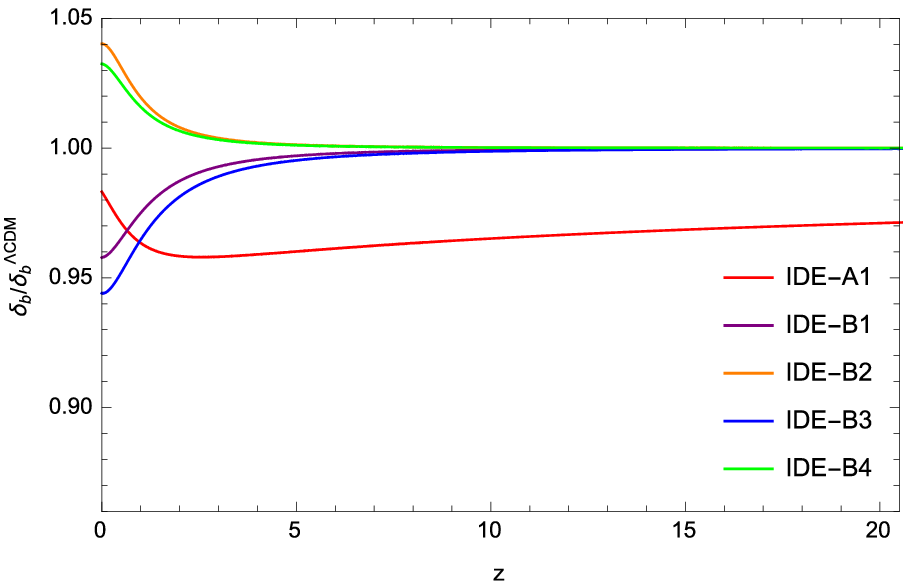}
}
\caption{\label{fig.H_deltab} Left: The ratio of Hubble parameter between different models and the fiducial $\Lambda$CDM model. Right: The ratio of linear perturbation to baryon density between different models and the fiducial $\Lambda$CDM model at scale $k = 0.034\mathrm{Mpc}^{-1}$. }
\end{figure*}

\begin{figure*}
\subfloat[]{
    \includegraphics[width=0.45\textwidth]{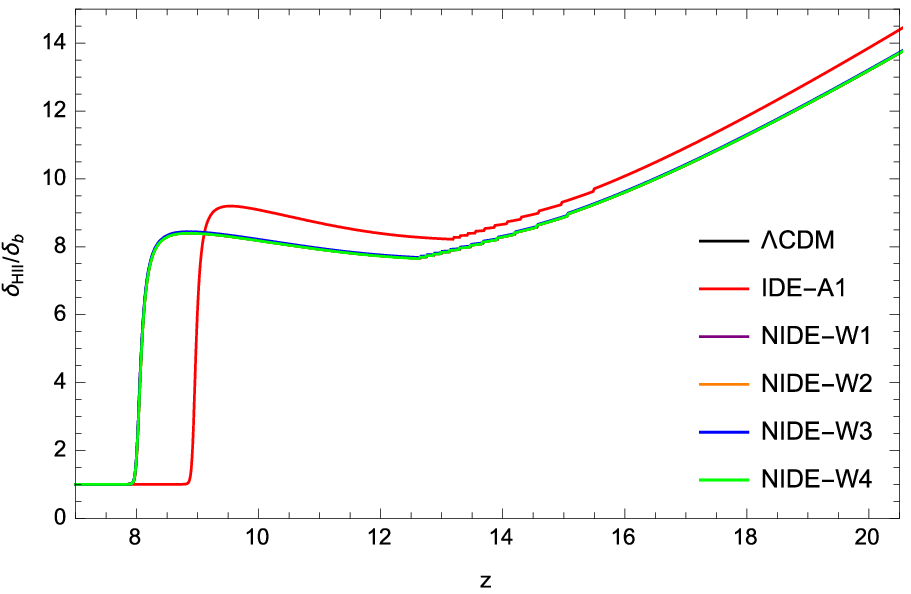}
}
\subfloat[]{
    \includegraphics[width=0.45\textwidth]{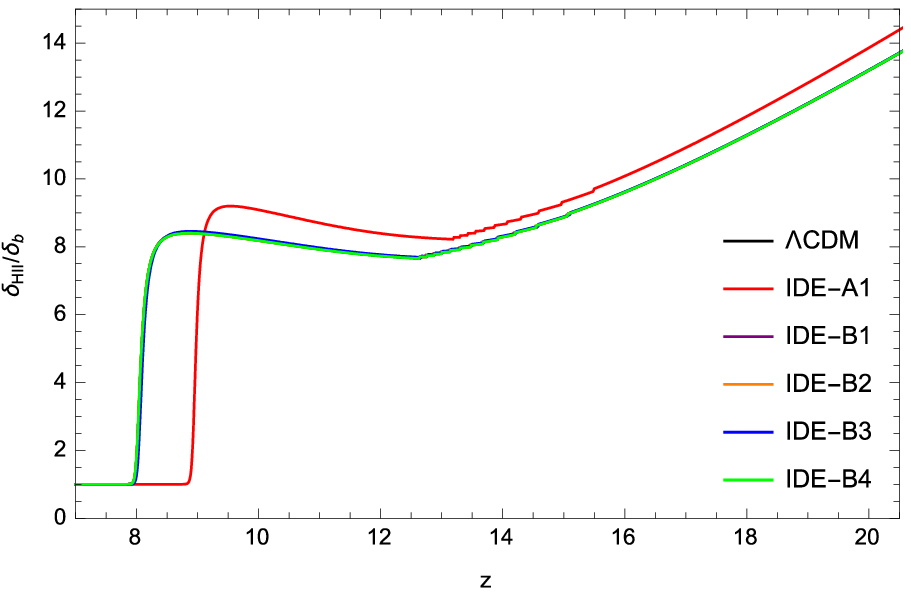}
}
\caption{\label{fig.bias_i} The bias between $\delta_{\mathrm{HII}}$ and $\delta_{\mathrm{b}}$ at scale $k=0.034\mathrm{Mpc}^{-1}$.}
\end{figure*}

In Fig.\ref{fig.f_HII} we  show the evolution of the ionized fraction in the models listed in Table.\ref{tab.parameters_w} and Table.\ref{tab.parameters_ide}. From Fig.\ref{fig.f_HII}(a) we find that DE equation of state has little effect on the reionization history of the universe. In Fig.\ref{fig.f_HII}(b), we also observe that the interaction between dark sectors in proportional to the energy density of DE is of little importance in the reionization history. In these models (NIDE-W and IDE-B), the evolution of $f_{\mathrm{HII}}$ is indistinguishable from that of the $\Lambda$CDM model. The little influence caused by NIDE-W and IDE-B models on the reionization history is not surprising, since the DE was negligible in the reionization era and the expansion and structure formation of the universe are not influenced much by DE and the interacting model IDE-B. In Fig.\ref{fig.H_deltab} it is clearly shown that the Hubble parameter and baryon overdensities in the NIDE-W and IDE-B models are almost the same as in $\Lambda$CDM model before the end of reionization. These background behaviors support that in models NIDE-W and IDE-B the reionization histories are almost the same as in $\Lambda$CDM model. In Fig.\ref{fig.bias_i}, we illustrate the bias of the fluctuations to the ionized fraction, $\delta_{\mathrm{HII}}$, with respect to the baryon overdensity $\delta_{\mathrm{b}}$ for $k=0.034\mathrm{Mpc}^{-1}$.  On other scales, the qualitative behaviors of perturbations are similar. Again we observe that the bias in NIDE-W and IDE-B models are almost indistinguishable from that of the $\Lambda$CDM model.

Choosing the interaction proportional to the energy density of DM,  it is clear in Fig.\ref{fig.f_HII} that the evolution of $f_{\mathrm{HII}}$ changes significantly. In order to avoid the instability in perturbations and keep DE energy density always positive, we choose $\xi_1 > 0$ and constant DE EoS $w<-1$ \cite{Wang2016}. With this kind of coupling between dark sectors, the reionization process is accelerated comparing to that in the $\Lambda$CDM model. In the fiducial $\Lambda$CDM model,  the reionization finished at $z \simeq 7.8$, while in our model IDE-A1, it finished at $z \simeq 8.8$. From Fig.\ref{fig.bias_i}.  We see that for IDE-A1 model(red line), the HII bias  evolves faster, which is in accordance with the average ionized fraction $f_{\mathrm{HII}}$. The large scale bias of the HII regions is always greater than unity during the reionization era, which means that the high density region tends to be more ionized than low density region-the topology of the reionization is inside-out. Then $\delta_{\mathrm{HII}}/\delta_{\mathrm{b}}$ drops to unity, indicating the end of reionization.

In IDE-A1 model the accelerated reionization is induced by the high Hubble parameter during the reionization era. The Thomson optical depth back to the same redshift, given the same ionized fraction, would be smaller if the expansion rate is enhanced. Therefore, to provide the same optical depth as the fiducial model, reionization must be faster and finishes earlier, and therefore the source emissivity should also be larger. We can see from Table.\ref{tab.parameters_ide} that the total number of ionizing photons  ($\zeta$) generated in IDE-A1 model is much higher than that of the fiducial $\Lambda$CDM model.

\begin{figure*}
\subfloat[]{
    \includegraphics[width=0.45\textwidth]{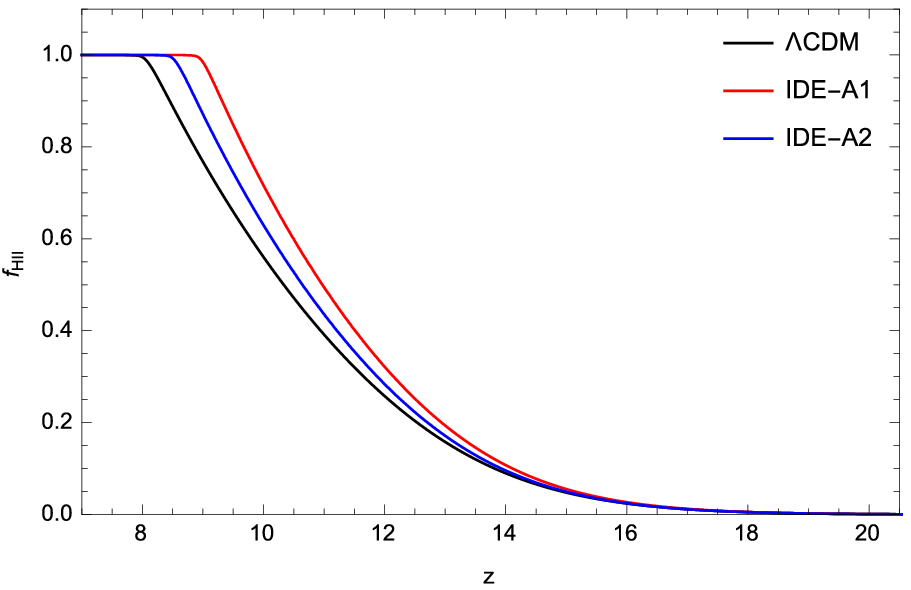}
}
\subfloat[]{
    \includegraphics[width=0.45\textwidth]{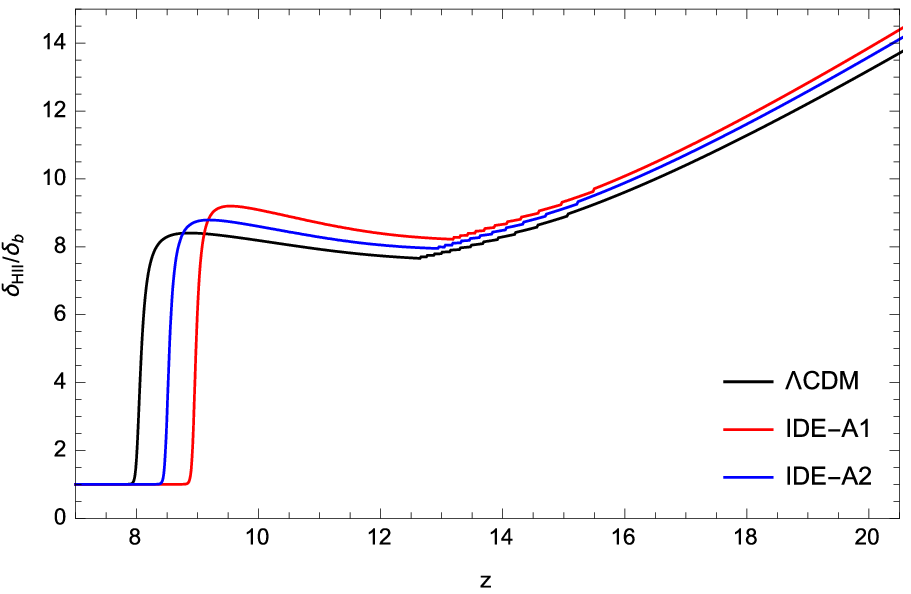}
}
\caption{\label{fig.fe_bias} (a) The evolution of ionized fraction in IDE-A models with different parameters and the fiducial $\Lambda$CDM model. (b) The bias between $\delta_{\mathrm{HII}}$ and $\delta_{\mathrm{b}}$ at scale $k=0.034\mathrm{Mpc}^{-1}$ in IDE-A models and the fiducial $\Lambda$CDM model. }
\end{figure*}

Considering that the expansion history and the structure growth can be altered much in IDE-A1 model,  in the following discussion we will focus on the interacting DE model with the interaction proportional to the energy density of DM. In Fig.\ref{fig.fe_bias} we exhibit the influence of model parameters, such as $w$ and $\xi_1$, on the ionized fraction and the bias of HII during the reionization era. We choose model parameters listed in Table.\ref{tab.parameters_ide}. We observe that with the increase of the coupling constant $\xi_1$ the HII bias shoots up more quickly before the time of percolation, so that the reionization is significantly speeded up, given the same Thomson optical depth. Thus, by studying the reionization history, one can break the degeneracy between DE equation of state $w$ and the coupling constant $\xi_{1}$, which generally exists in the low redshift observations as well as CMB\cite{Costa2013, He2010, He2009a, He2009, Xu2011, Xu2013}. Instead, $\xi_1$ is now degenerate with the emissivity($\zeta$) or optical depth($\tau_{\mathrm{T}}$).

\begin{figure*}
\subfloat[]{
    \includegraphics[width=0.3\textwidth]{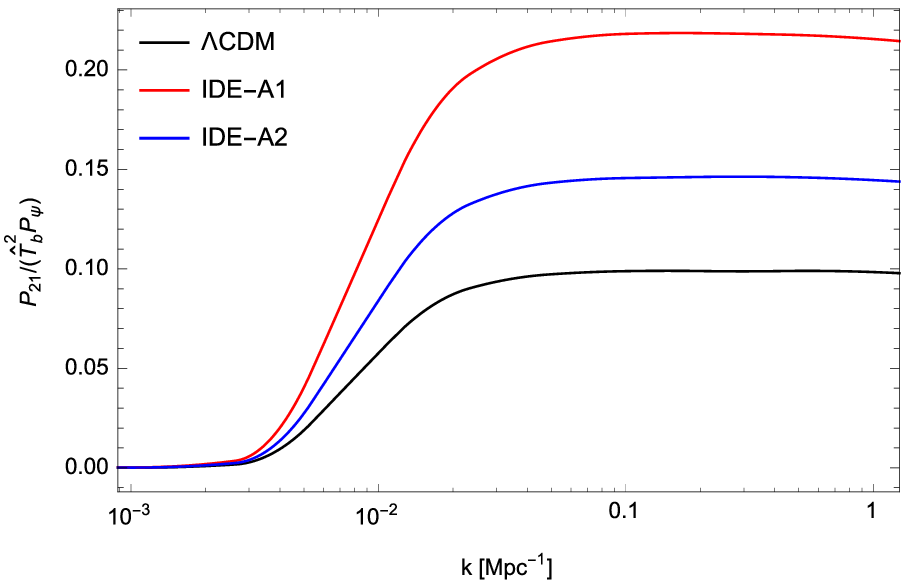}
}
\subfloat[]{
    \includegraphics[width=0.3\textwidth]{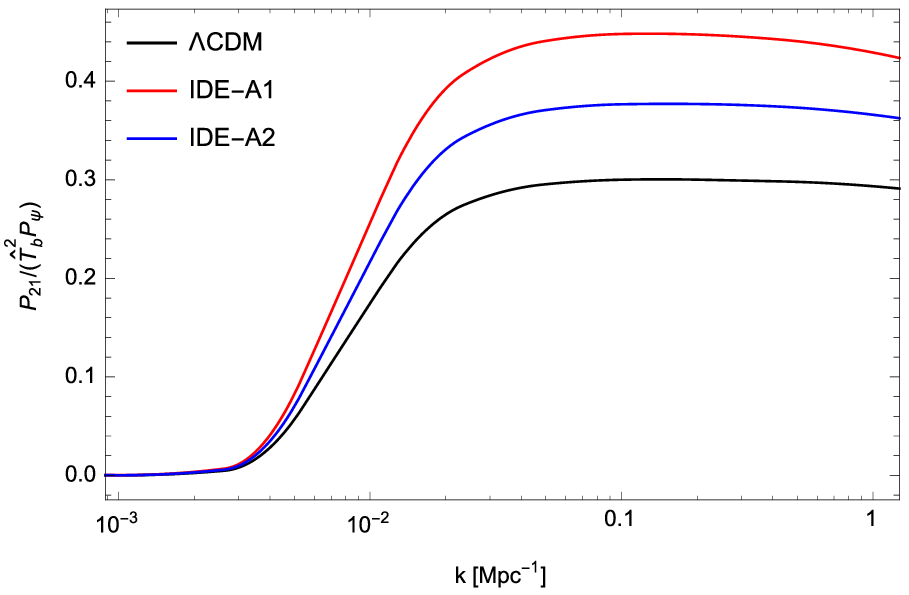}
}
\subfloat[]{
    \includegraphics[width=0.3\textwidth]{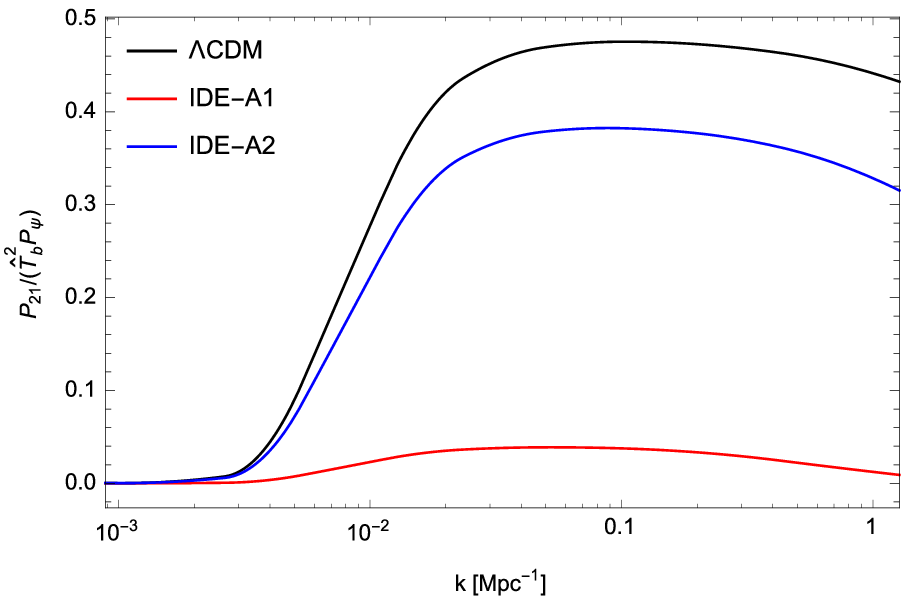}
}
\caption{\label{fig.P21_z} The 21cm power spectrum for IDE-A and $\Lambda$CDM model at (a) $z=11$, (b) $z=10$ and (c) $z=9$. }
\end{figure*}

\begin{figure*}[tp]
\subfloat[]{
    \includegraphics[width=0.3\textwidth]{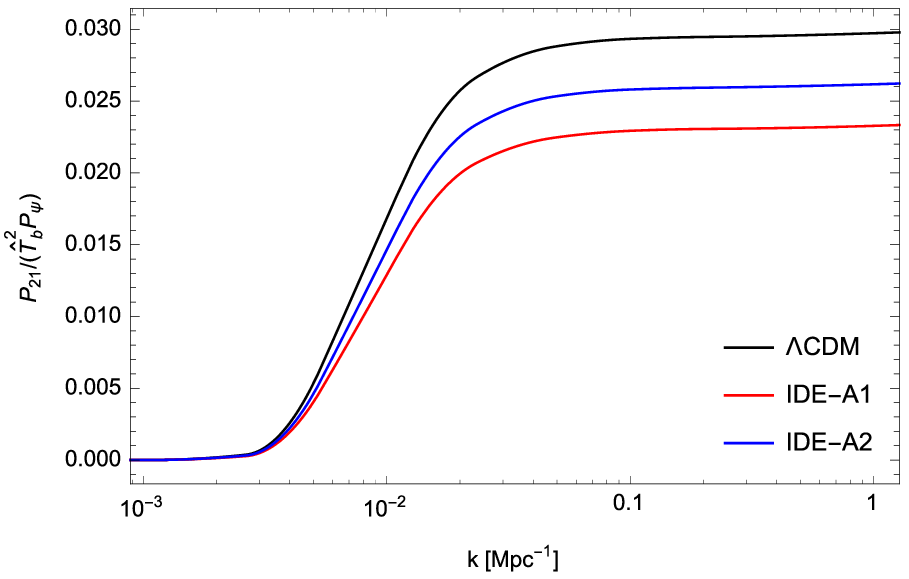}
}
\subfloat[]{
    \includegraphics[width=0.3\textwidth]{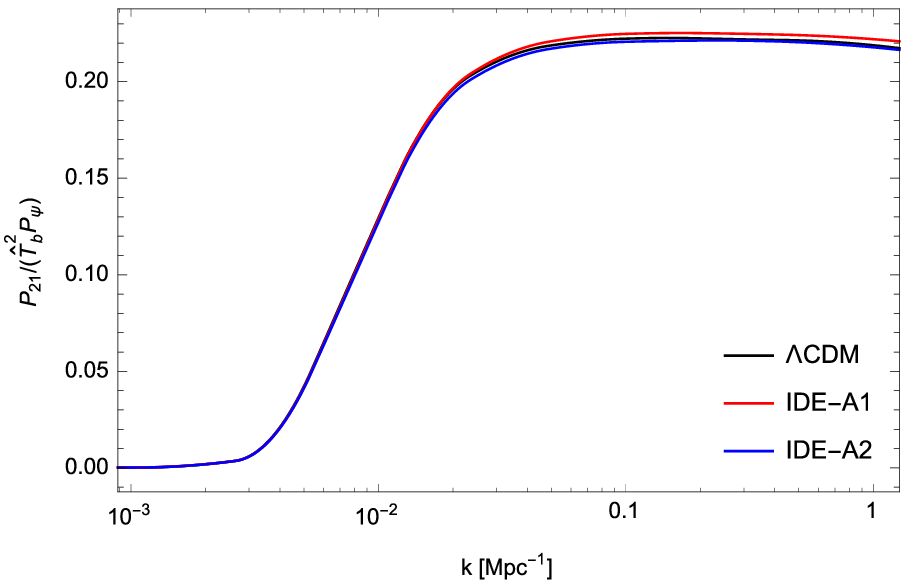}
}
\subfloat[]{
    \includegraphics[width=0.3\textwidth]{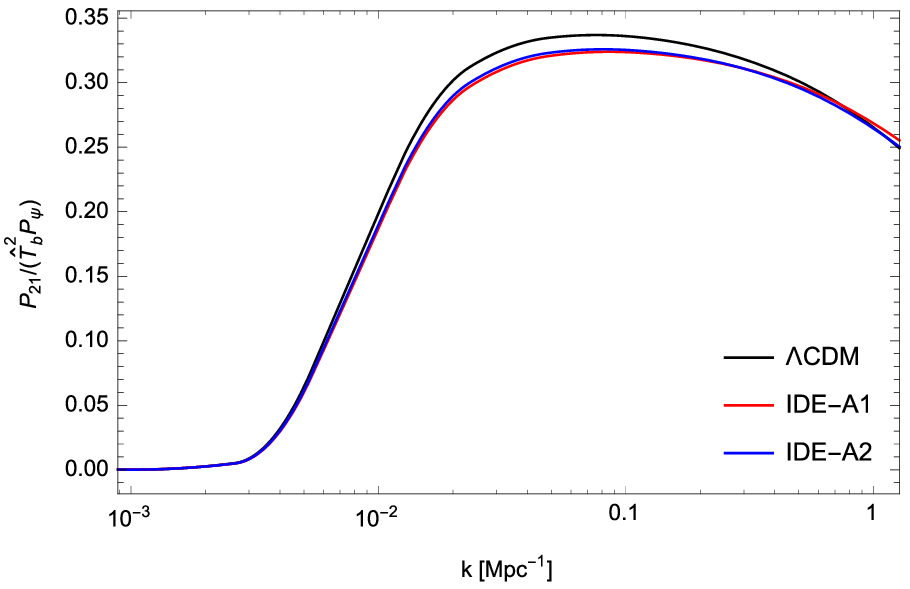}
}
\caption{\label{fig.P21_xHII} The 21cm power spectrum for IDE-A and $\Lambda$CDM model at (a) $f_{\mathrm{HII}}=0.1$, (b) $f_{\mathrm{HII}}=0.5$ and (c) $f_{\mathrm{HII}}=0.9$. }
\end{figure*}

\begin{figure*}[tp]
\subfloat[]{
    \includegraphics[width=0.45\textwidth]{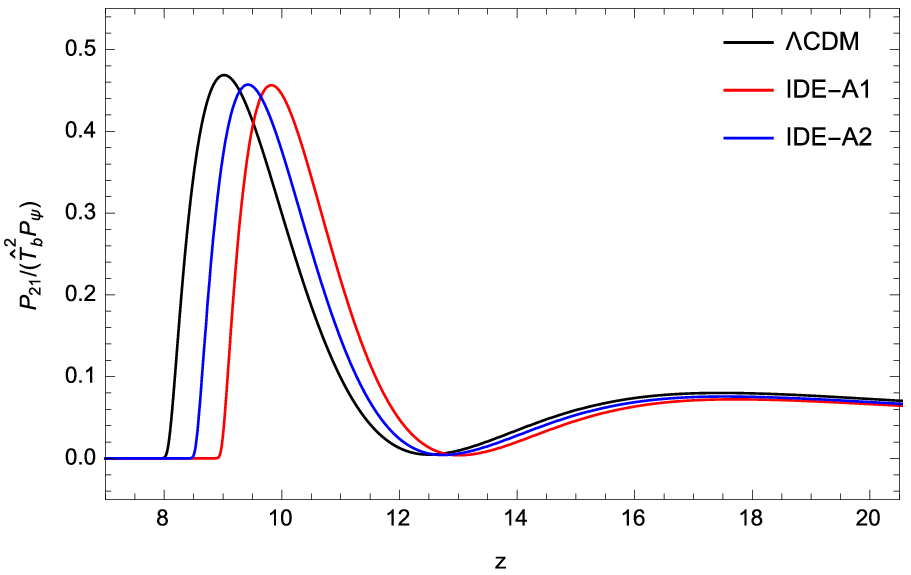}
}
\subfloat[]{
    \includegraphics[width=0.45\textwidth]{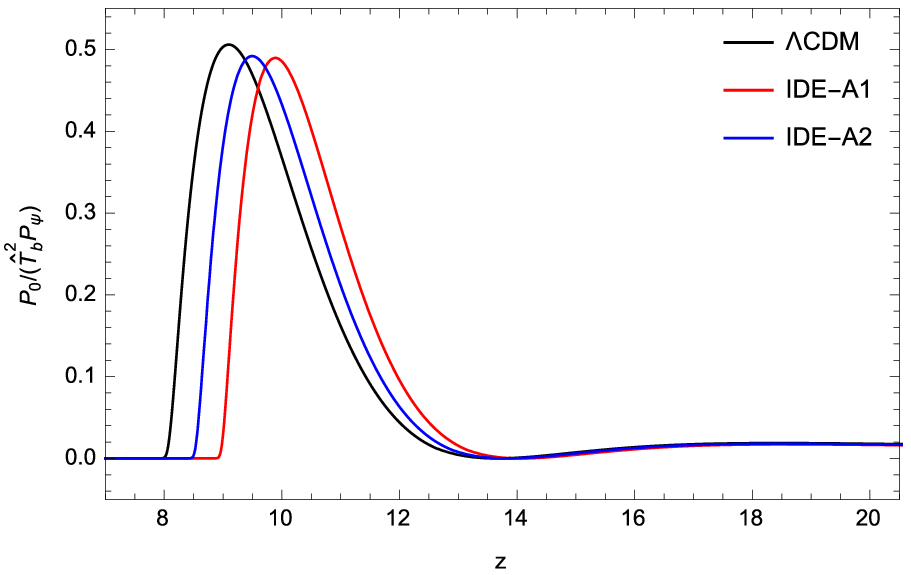}
}\\
\subfloat[]{
    \includegraphics[width=0.45\textwidth]{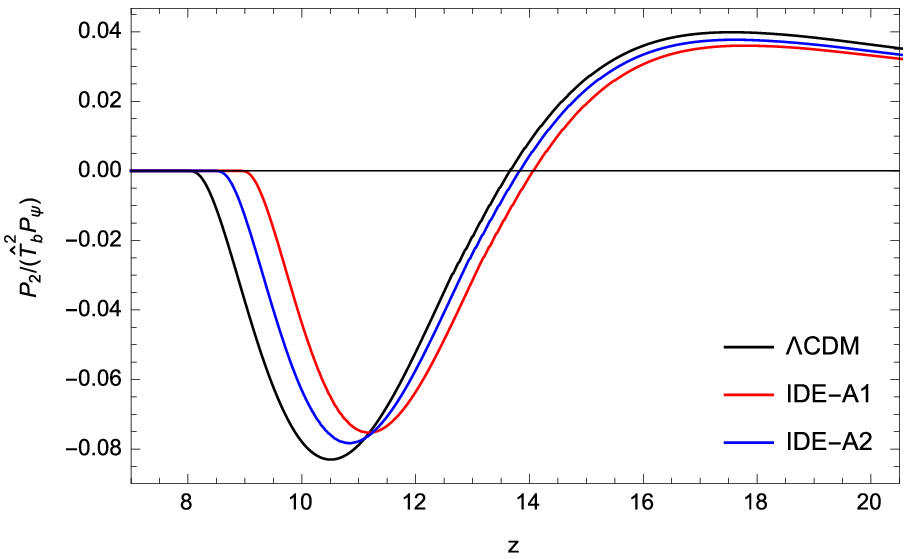}
}
\subfloat[]{
    \includegraphics[width=0.45\textwidth]{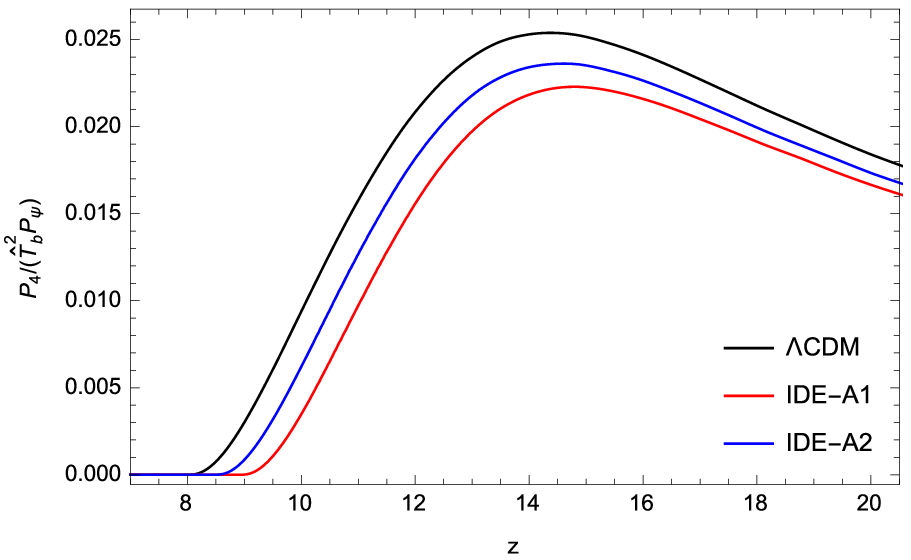}
}
\caption{\label{fig.P21_k} The 21cm power spectrum $P_{21}(k)$ and its components for IDE-A and $\Lambda$CDM model at $k=0.034\mathrm{Mpc}^{-1}$. }
\end{figure*}

With numerical values of $f_{\mathrm{HI}}$ and $\delta_{\mathrm{HII}}$ in the presence of interaction between dark sectors, we can now calculate the 21cm power spectrum in the interacting DM-DE models. Since only the model with interaction between dark sectors proportional to the energy density of DM shows significant effect in the reionization era, we will concentrate on the IDE-A models below. We plot the 21cm power spectrum $p_{21}(k) \equiv P_{21}(k)/(\hat{T}^2_{\mathrm{b}}P_{\Psi}(k))$ in two IDE-A models and the $\Lambda$CDM model in Fig.\ref{fig.P21_z} at $z=9,10,11$, respectively. $P_{\Psi}(k)$ is the primordial fluctuation power spectrum and $\hat{T}_{\mathrm{b}} \equiv 3hc^3A_{10}\bar{n}_{\mathrm{b}}^0 (1-f_{\mathrm{He}}) / (32\pi k_{\mathrm{B}}\nu_0^2 H_0)$, where $\bar{n}_{\mathrm{b}}^0$ is the baryon number density at present. The shape of $p_{21}(k)$ for three models are similar. It is almost scale independent for subhorizon modes at earlier times, and later on the power at small scales becomes slightly lower than that on large scales. At $z=11$ and $z=10$, we find that the amplitude of $p_{21}(k)$ in the IDE-A1 model is the highest and the $\Lambda$CDM model is the lowest. When $z=9$, however, the relative amplitudes among the models are reversed. This is attributed to that in the IDE-A models, reionization finishes earlier, and $p_{21}(k)$ drops rapidly with increasing $f_{\mathrm{HII}}$. If calculate the 21cm power spectra at the same $f_{\mathrm{HII}}$ instead of the same $z$, as shown in Fig.\ref{fig.P21_xHII}, we find that the power sepctrum in $\Lambda$CDM model becomes closer to that in IDE-A models, or even exceeds the latter. This can be seen more clearly in Fig.\ref{fig.P21_k}, in which we plotted the evolution of $P_{21}(z)$ as well as its components $P_0(z)$, $P_2(z)$ and $P_4(z)$ for $k=0.034\mathrm{Mpc}^{-1}$, assuming $(\hat{\mathbf{n}} \cdot \hat{\mathbf{k}})^2 = 1$. In all models, $P_{21}(z)$ exhibits the similar behavior. It first grows with time. At this stage, the ionization bias $\delta_{\mathrm{HII}}/\delta_{\mathrm{b}}$ evolves slowly, the growth of power is mainly due to the growth of baryon overdensity $\delta_{\mathrm{b}}$. After the majority of hydrogen atoms are ionized, the bias quickly drops and approaches to unity, which, together with the decrease of neutral hydrogen, leads to the suppression of 21cm power. Enhancing the coupling strength, $P_{21}$ grows, peaks and drops earlier, and the amplitude of the peak becomes lower.

Among the components of $P_{21}(k)$, $P_4(k)$ is of special interest, because it only depends on $P_{\mathrm{bb}}$ and do not involve the fluctuation to ionized fraction, thus can be used to probe the growth of structure during reionization era without relying on the detailed knowledge of reionization itself. Yet it is not entirely independent of reionization model. From \eqref{eq.P_4}, $P_4(k)$ is still modulated by the spatially averaged ionized fraction $f_{\mathrm{HII}} = 1 - f_{\mathrm{HI}}$. Our calculations suggest that $P_4(k)$ is always subdominant comparing with other components except at the beginning of reionization, which is a result of inside-out reionization, in which $\delta_{\mathrm{HII}}/\delta_{\mathrm{b}}$ is always greater than $1$ and $P_{\mathrm{ii}}$ and $P_{\mathrm{ib}}$ are greater than $P_{\mathrm{bb}}$. But this would not hinder the measurement of $P_4(k)$. One can isolate $P_4(k)$ from overall 21cm radiation signals through its angular dependence-$P_4(k)$ in proportional to $(\hat{\mathbf{n}} \cdot \hat{\mathbf{k}})^4$, which render it distinguishable from other components. Other than $P_0(k)$ and $P_2(k)$, $P_4(k)$ in the IDE-A models never exceeds that in the $\Lambda$CDM model. This is actually a direct consequence of that $P_4(k)$ only depends on $f_{\mathrm{HII}}$ but not on $\delta_{\mathrm{HII}}$. In IDE-A models, hydrogen gas is ionized faster, hence both $f_{\mathrm{HII}}$ and $\delta_{\mathrm{HII}}$ are larger than those in $\Lambda$CDM model at given redshift. $P_4(k)$ is then suppressed by low neutral hydrogen density. On the other hand, the other two components are enhanced due to high $\delta_{\mathrm{HII}}$ and can be larger than those of $\Lambda$CDM model.

\subsection{Best-fit models}

In Ref \cite{Costa2017}, phenomenological interacting DM-DE models were constrained by employing the recent cosmological data including the cosmic microwave background radiation anisotropies from Planck 2015, Type Ia supernovae, baryon acoustic oscillations, the Hubble constant and redshift-space distortion datasets. In this subsection, we will use numerical fitting results to see what effects these best-fit models can have on the reionization history. Table \ref{tab.models} shows the phenomenological models we are going to investigate. Here we will choose the results in Ref \cite{Costa2017} by using the combination of Planck2015, BAO, SNIa and $\mathrm{H}_0$ data sets. The constrained model parameters are listed in Table \ref{tab.model_parameters}. In the computation, we also tune $\zeta$ for each model in order to let $\tau_{\mathrm{T}}$ in consistent with the constrained value from observational data.

\begin{table*}
\caption{\label{tab.models} Phenomenological interaction dark energy models.}
\begin{tabular}{p{60pt}p{80pt}p{80pt}p{80pt}}
    \toprule
     Model & $Q$ & $w$ & Constraints \\
    \hline
    I & $3\xi_{2}H\rho_{d}$ & $-1<w<0$ & $\xi_2<0$ \\
    II & $3\xi_{2}H\rho_{d}$ & $w<-1$ & $0<\xi_2<-2w\Omega_c$ \\
    III & $3\xi_{1}H\rho_{c}$ & $w<-1$ & $0<\xi_1<-w/4$ \\
    IV & $3\xi H(\rho_{c}+\rho_{d})$ & $w<-1$ & $0<\xi_1<-w/4$ \\
    \lasthline
\end{tabular}
\end{table*}

\begin{table*}
\caption{\label{tab.model_parameters} Model parameters}
\begin{tabular}{p{45pt}p{40pt}p{50pt}p{50pt}p{50pt}p{50pt}p{50pt}p{50pt}p{40pt}}
    \toprule
     Model & $H_0$ & $\Omega_bh^2$ & $\Omega_ch^2$ & $\tau_{\mathrm{T}}$ & $w$ & $\xi_1$ & $\xi_2$ & $\zeta$\\
    \hline
    I & 68.18 & 0.02223 & 0.0792 & 0.08204 & -0.9191 & - & -0.1107 & 88.9\\
    II & 68.35 & 0.02224 & 0.1351 & 0.081 & -1.088 & - & 0.05219 & 90.3\\
    III & 68.91 & 0.02228 & 0.1216 & 0.07728 & -1.104 & 0.0007127 & - & 79.7\\
    IV & 68.88 & 0.02228 & 0.1218 & 0.07709 & -1.105 & 0.000735 & 0.000735 & 79.6\\
    \lasthline
\end{tabular}
\end{table*}

\begin{figure*}
\subfloat[]{
    \includegraphics[width=0.45\textwidth]{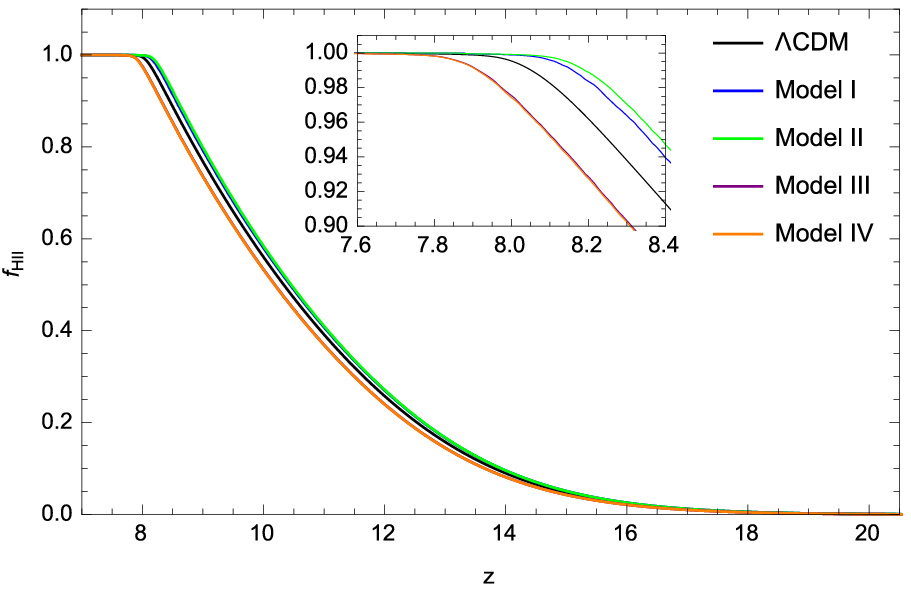}
}
\subfloat[]{
    \includegraphics[width=0.45\textwidth]{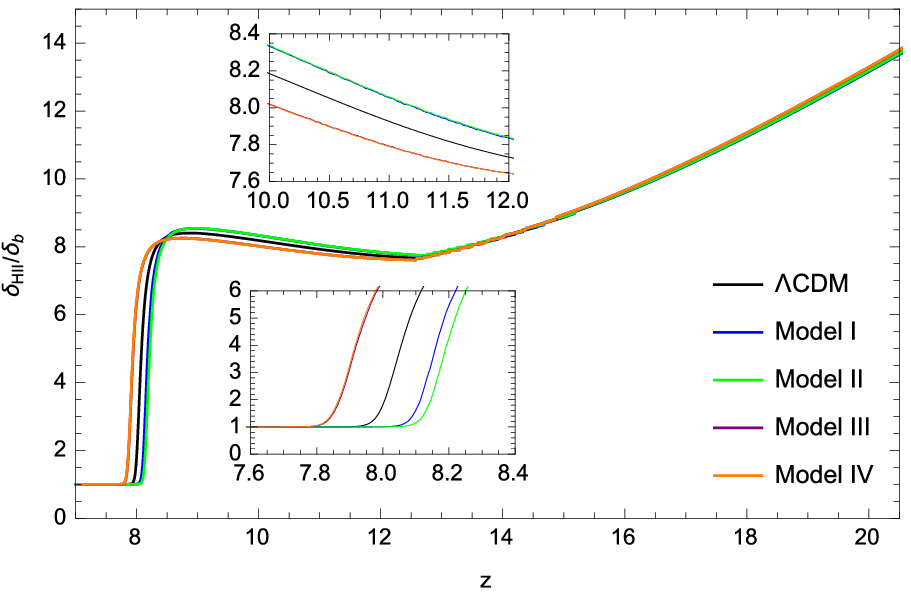}
}
\caption{\label{fig.fe_bias_1234} (a) The evolution of ionized fraction in Model I, II, III, IV and the fiducial $\Lambda$CDM model. (b) The bias between $\delta_{\mathrm{HII}}$ and $\delta_{\mathrm{b}}$ at scale $k=0.034\mathrm{Mpc}^{-1}$ in Model I, II, III, IV and the fiducial $\Lambda$CDM model. }
\end{figure*}

Fig.\ref{fig.fe_bias_1234}(a) shows the evolution of the ionized fraction in Model I, II, III and IV. We can see that although model parameters are close to $\Lambda$CDM model, the evolution of $f_{\mathrm{HII}}$ can still be distinguished from the $\Lambda$CDM model. We find that in Model I and II the reionization process is accelerated comparing to that in $\Lambda$CDM model, whereas for Model III and IV the reionization process slows down. The effect of Model III and IV in the reionization era is different from the qualitative discussion for model IDE-A1 in the previous subsection, which is caused by the small change in the optical depth value. Here we use the optical depth value from the best fitting result from observations. During the calculation we find that such small difference in the optical depth brings important influence on the reionization process. Fig.\ref{fig.fe_bias_1234}(b) shows the bias between $\delta_{\mathrm{HII}}$ and $\delta_{\mathrm{b}}$ at scale $k=0.034\mathrm{Mpc}^{-1}$, which is in accordance with the average ionized fraction $f_{\mathrm{HII}}$ as shown in Fig.\ref{fig.fe_bias_1234}(a). Model III and Model IV lines degenerate. This is due to the fact that the constrained parameters of these two models are very similar to each other as listed in Table \ref{tab.model_parameters}.

\begin{figure*}
\subfloat[]{
    \includegraphics[width=0.3\textwidth]{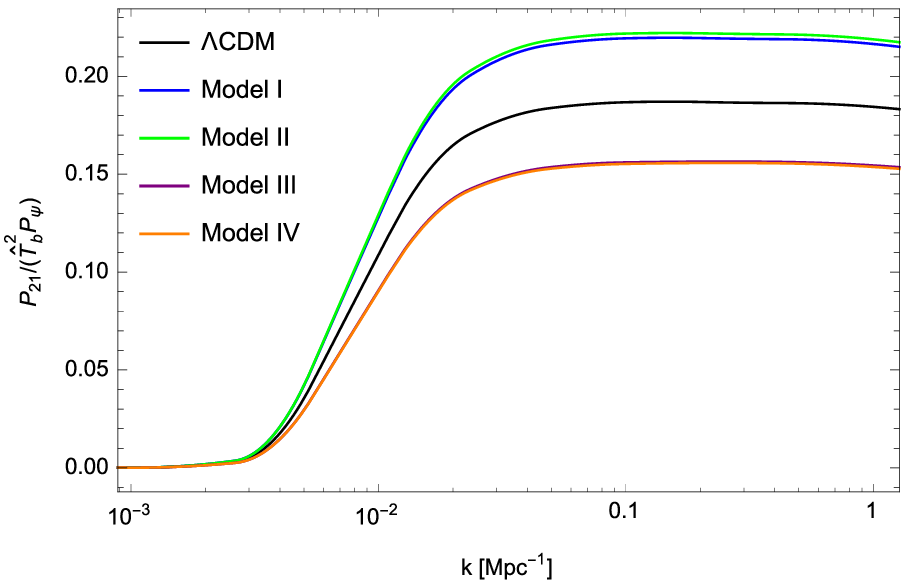}
}
\subfloat[]{
    \includegraphics[width=0.3\textwidth]{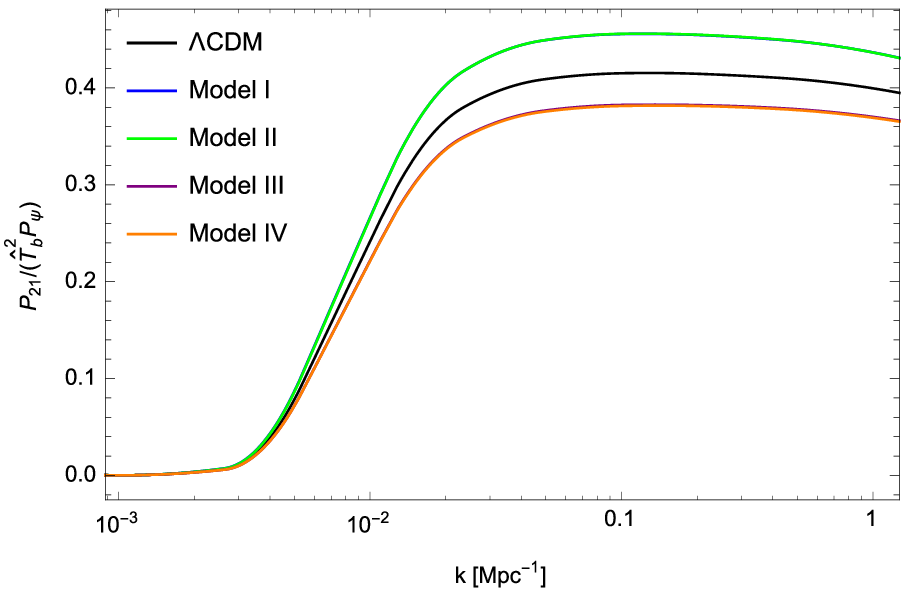}
}
\subfloat[]{
    \includegraphics[width=0.3\textwidth]{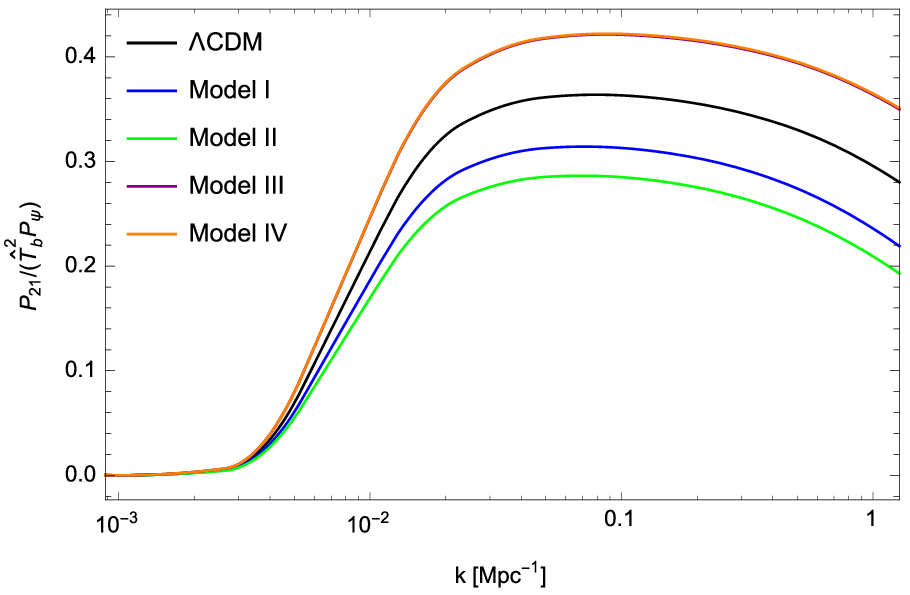}
}
\caption{\label{fig.P21_z_1234} The 21cm power spectrum for Model I, II, III, IV and $\Lambda$CDM model at (a) $z=10.5$, (b) $z=9.5$ and (c) $z=8.5$. }
\end{figure*}

\begin{figure*}[tp]
\subfloat[]{
    \includegraphics[width=0.3\textwidth]{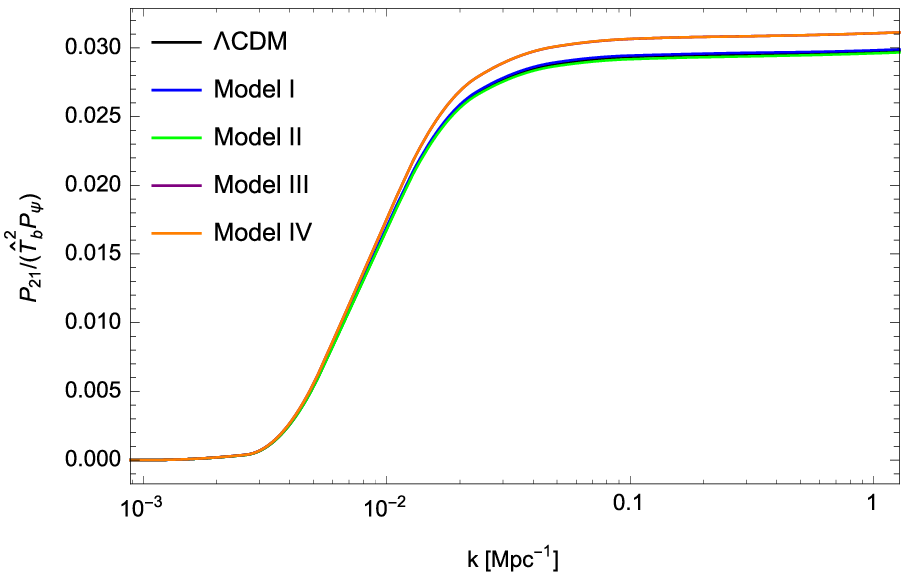}
}
\subfloat[]{
    \includegraphics[width=0.3\textwidth]{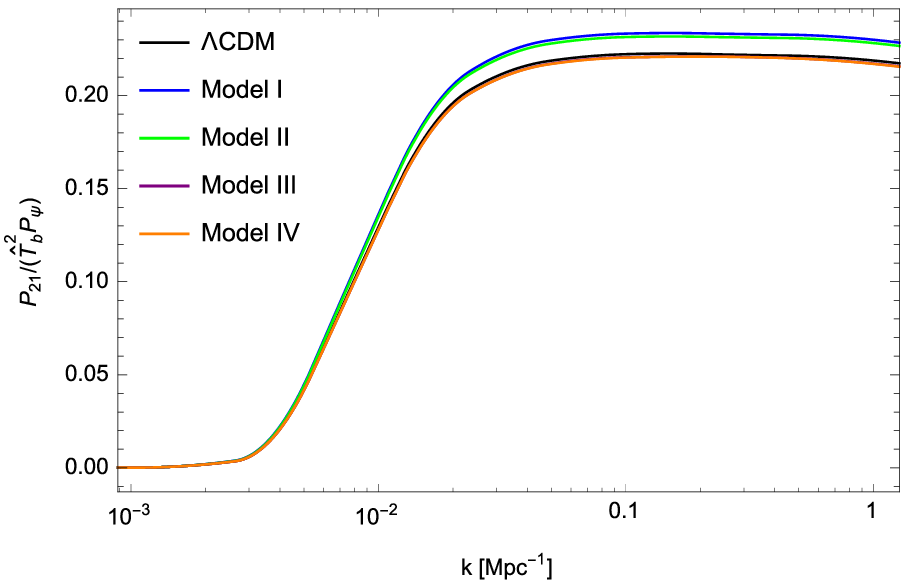}
}
\subfloat[]{
    \includegraphics[width=0.3\textwidth]{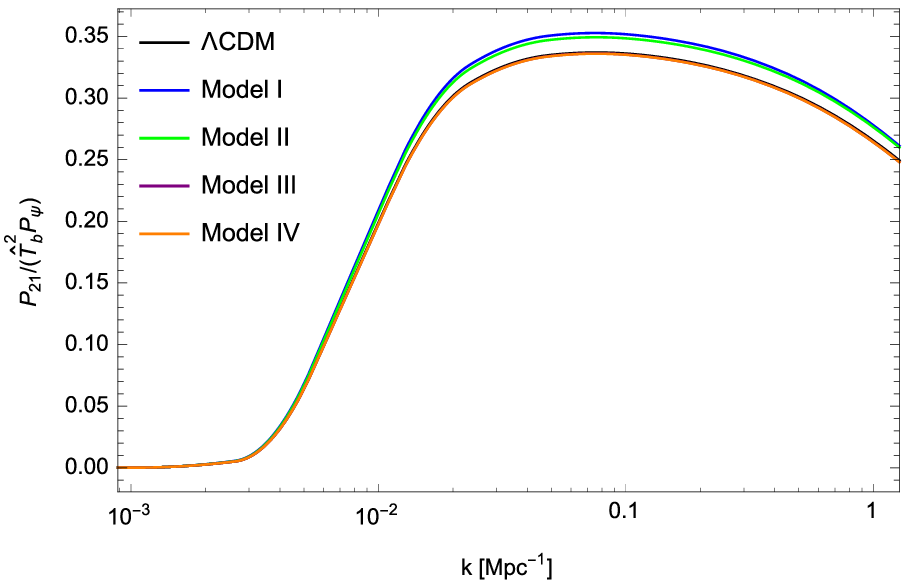}
}
\caption{\label{fig.P21_xHII_1234} The 21cm power spectrum for Model I, II, III, IV and $\Lambda$CDM model at (a) $f_{\mathrm{HII}}=0.1$, (b) $f_{\mathrm{HII}}=0.5$ and (c) $f_{\mathrm{HII}}=0.9$. }
\end{figure*}

\begin{figure*}[tp]
\subfloat[]{
    \includegraphics[width=0.45\textwidth]{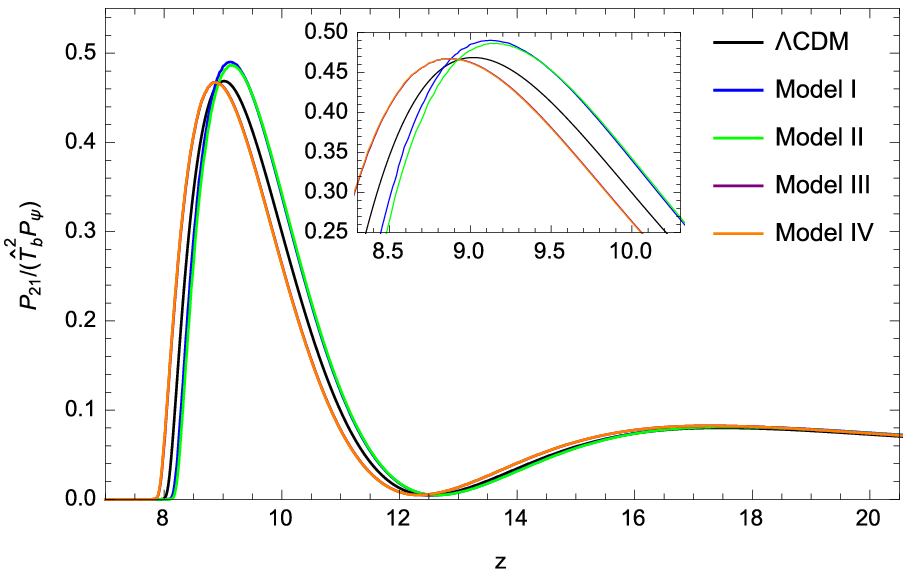}
}
\subfloat[]{
    \includegraphics[width=0.45\textwidth]{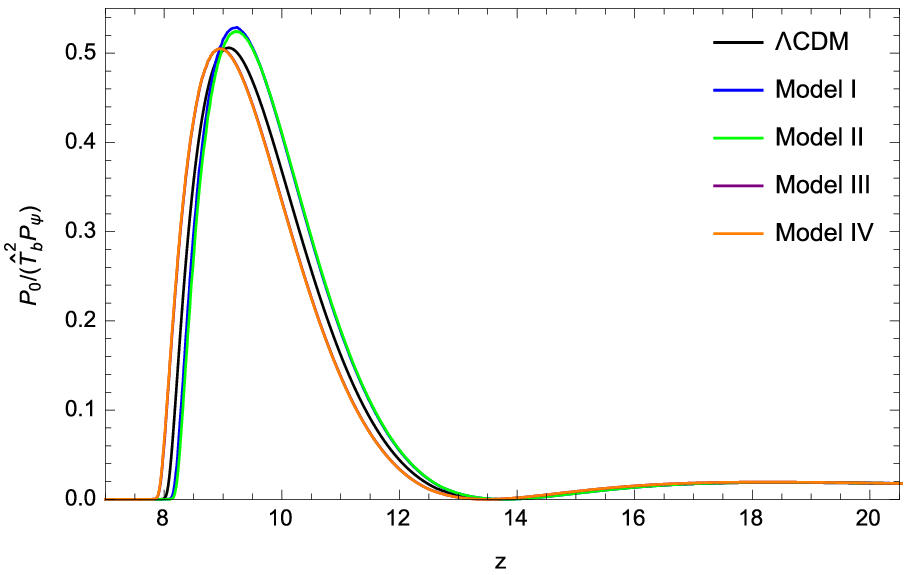}
}\\
\subfloat[]{
    \includegraphics[width=0.45\textwidth]{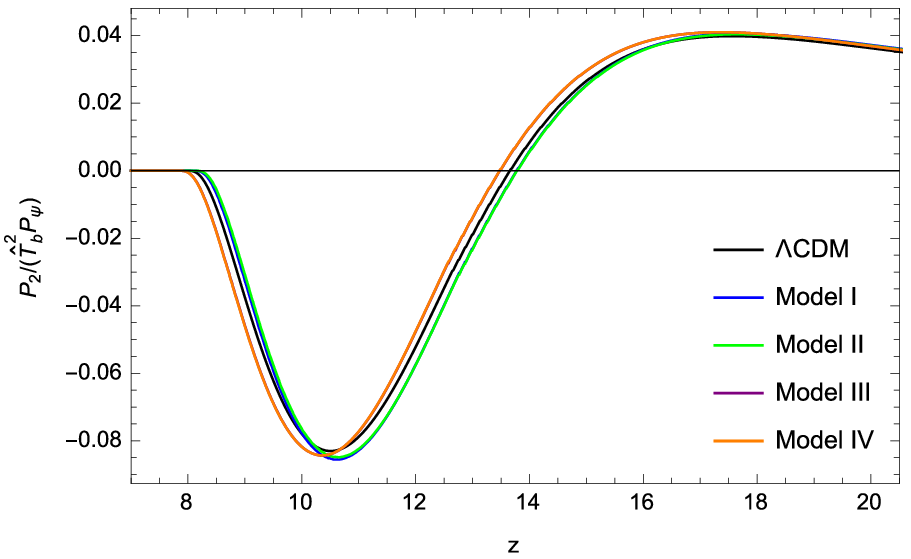}
}
\subfloat[]{
    \includegraphics[width=0.45\textwidth]{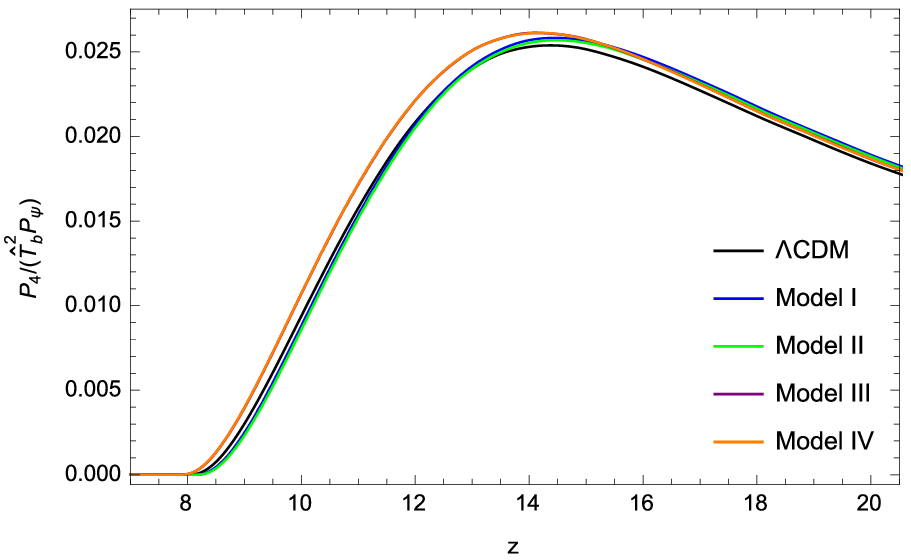}
}
\caption{\label{fig.P21_k_1234} The 21cm power spectrum $P_{21}(k)$ and its components for Model I, II, III, IV and $\Lambda$CDM model at $k=0.034\mathrm{Mpc}^{-1}$. }
\end{figure*}

Fig.\ref{fig.P21_z_1234} shows the 21cm power spectrum $p_{21}(k) \equiv P_{21}(k)/(\hat{T}_{\mathrm{b}}P_{\Psi}(k))$ at different redshift $z=8.5,9.5,10.5$ in Model I, II, III, IV and the $\Lambda$CDM model. Since the reionization process finishes earlier in Model I and II than that in the $\Lambda$CDM model, the corresponding $p_{21}(k)$ emission spectrum drops quicker from $z=10.5$ to $z=8.5$ as shown in Fig.\ref{fig.P21_z_1234}. In Model III and IV, however, the effect is opposite. Fig.\ref{fig.P21_xHII_1234} shows the 21cm power spectra at different ionized fraction $f_{\mathrm{HII}}=0.1,0.5,0.9$. We can see that the power spectra in these models are very close to each other. To see more clearly, we plot the evolution of $P_{21}(z)$ as well as its components $P_0(z)$, $P_2(z)$ and $P_4(z)$ for $k=0.034\mathrm{Mpc}^{-1}$ in Fig.\ref{fig.P21_k_1234}. We find that in Model I and II, $P_{21}$ grows, peaks and drops earlier, and in Model III and IV, this process becomes slower comparing to the $\Lambda$CDM model. The amplitude of the peak in the $\Lambda$CDM model is lower than the other four interacting models, among which the amplitudes in Model I and II are higher than that in Model III and IV.

\section{Conclusions \label{sec.conclusions}}

In this work, we investigate reionization in an IDE scenario. We assume a phenomenological parametrization of the interaction between dark matter and dark energy, and study its influence on reionization.  To examine qualitative influence on the reionization caused by different models, we first fix the optical depth as that of the $\Lambda$CDM model. We find that changing the equation of state of DE or adopting the interaction between dark sectors in proportional to DE energy density would not alter reionization process. However, when the interaction proportional to DM energy density, it will significantly affect the reionization. Comparing with $\Lambda$CDM model, EoR needs to end earlier if Thomson optical depth remains the same. This implies stronger source emissivity of ionizing photons. Alternatively, if the source emissivity is fixed, the expected optical depth will change with the coupling strength of the interaction. On the perturbation level, we find that the linear fluctuations to the ionized fraction in IDE models behave similarly as in the $\Lambda$CDM model. For NIDE-W and IDE-B models, the evolution of HII bias is indistinguishable from $\Lambda$CDM model, while for IDE-A models, $\delta_{\mathrm{HII}}(k)$ evolves faster given the same optical depth.

To investigate the signature of interaction between DM and DE during EoR, we compute the 3D power spectrum of redshifted 21 cm radiation in IDE-A models. Our theoretical predictions of $P_{21}(k)$ for IDE-A share the similar shape to that of the $\Lambda$CDM model, however there is a prominent time shift between them. The time shift depends on both the coupling strength and ionizing source emissivity. It is degenerated with the optical depth in the 21 cm signals. Possible detection of the 21 cm signals from redshift range before and after EoR can in principle break such degeneracy. We will explore this in the future work.

We examine the reionization history in the best-fit interacting DM-DE models. Each model has the best constrained optical depth, which slightly deviates from the $\Lambda$CDM model value. Comparing with the qualitative study, we find that such small difference in the optical depth brings important influence. We find that the evolution of HII bias in these models can be distinguished from the $\Lambda$CDM model. The reionization progress in Model I and II finishes earlier than that in $\Lambda$CDM model. While for the other two models, the effect is opposite. This is an interesting phenomenon, since it tells us that behavious of different cosmological models may reflect on the reionization history.

It is worth noting that the validity of linear perturbation theory on large scales needs further investigation, as small scale fluctuations can contribute to large scale modes through mode-mode couplings due to the presence of the nonlinear terms in the equations of the ionization balance and radiative transfer. However, for the purpose of studying the effect of DM-DE interaction on reionization qualitatively, the linear perturbation calculation should be sufficient.

Our work has only considered reionization by a soft source spectrum (dominated by UV photons from stars). It is argued recently that AGN may play a significant role in re-ionizing the universe \cite{Kulkarni2017}, suggesting that our understanding of the first generation of luminous sources is still uncertain. Therefore, to constrain the DM-DE intereaction theories with the future 21cm observations during the epoch of reionization, we need to further increase the dimension of the parameter space in our calculation. Our linear perturbation calculation provides a convenient tool for this type of purposes. We will study more complicated reionization scenarios in the future.

In \cite{Bowman2018} the 21cm signal in the form of the absorption line against the CMB blackbody spectrum was obtained, where the spin temperature was lower than the CMB temperature. It was argued that a larger positive coupling between dark sectors can present a clearer difference in the intensity of the 21cm signal relative to the CMB temperature \cite{Costa2018}. In the absorption spectrum  the interaction can show up no matter it is proportional to the energy density of DE or DM. Here in the reionization discussion, adopting the best constrained model parameters for interacting DE models, we find that the effects shown in the evolution of ionized fraction and the 21cm emission spectra are different in various interacting models and can be distinguished from $\Lambda$CDM model. This is interesting, since it tells us that 21cm emission spectrum can be helpful for us to investigate different cosmological models.

\begin{acknowledgments}
X. X. is supported by the South African Research Chairs Initiative of the Department
of Science and Technology and National Research Foundation of South Africa as well as the Competitive Programme for Rated Researchers (Grant Number 91552). J. Z. is supported by the NSFC grants (11433001 and 11673016), the National Key Basic Research Program of China (2015CB857001), and a grant (No.11DZ2260700) from the Office of Science and Technology in Shanghai Municipal Government. B. W acknowledges financial support from National Basic Research Program of China (973 Program 2013CB834900 ) and National Natural Science Foundation of China (No. 11375113). BY acknowledges the support of the Hundred Talents (Young Talents) program from the CAS and the NSFC-CAS joint fund for space scientic satellites No. U1738125. We would like to acknowledge helpful discussions with Elcio Abdalla and Yidong Xu.

\end{acknowledgments}

\begin{appendices}
\section*{Appendix}

We understand that the semi-analytical model we adopted in the work is a rough approximation to the complex physics of the reionizaiton history. There are uncertainties in the model parameters which might give an effect on the final results. To clarify the reliability of this semi-analytical model, we used a semi-numerical modeling tool--21cmFAST \cite{Mesinger2011}--to simulate the cosmological 21cm signal, and did comparison with our semi-analytical results.

\begin{figure}[h]
\subfloat[]{
    \includegraphics[width=0.45\textwidth]{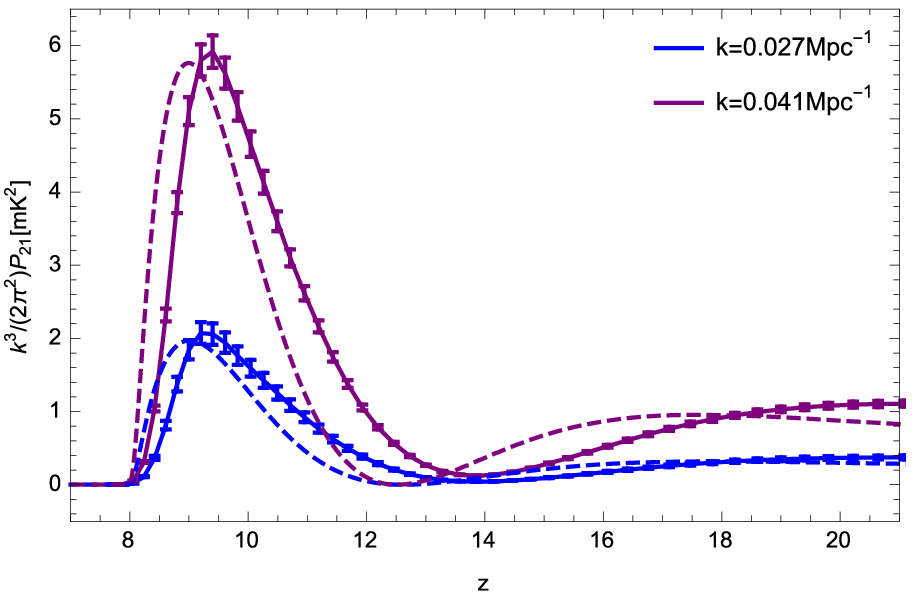}
}
\subfloat[]{
    \includegraphics[width=0.45\textwidth]{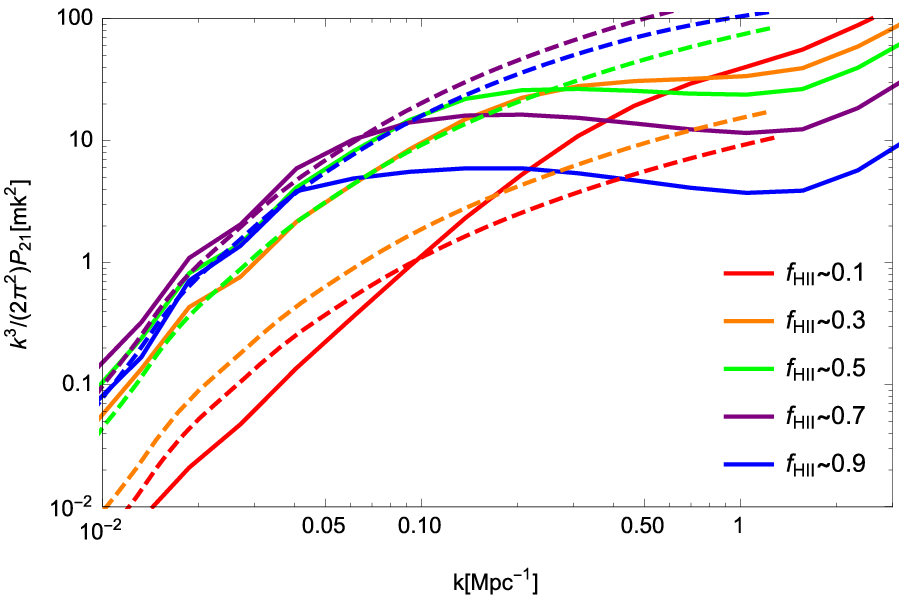}
}
\caption{\label{p21} Comparisons of 21cm power spectra obtained from the semi-analytical model (dashed lines) and 21cm FAST (solid lines) in the standard $\Lambda$CDM model.}
\end{figure}

For the standard $\Lambda$CDM model, we presented the 21cm power spectra $k^3/(2\pi^2)P_{21}(k)$[mK$^2$] at $k=0.017$ Mpc$^{-1}$, 0.041 Mpc$^{-1}$ obtained from our semi-analytical treatment and from the 21cmFAST with same cosmology and consistent reionization history in Fig. \ref{p21}(a). We can see that the 21cm power spectra exhibit the consistent behavior in semi-analytical approach and simulations. The error bars on 21cmFAST data represent  the statistical variance due to limited number of independent $k$ modes measured at the relevant  $k$  bins. In Fig. \ref{p21} (b) we show the 21cm power spectrum at different redshift $z$ or ionized fraction $f_{\text{HII}}$. Although semi-analytical results shown in dashed lines do not match exactly with numerical results exhibited in solid lines, when $k<0.1Mpc^{-1}$ they have similar shapes, and their time evolutions are also similar. In our work, we investigated the effect of the interaction between dark sectors on the reionization process and mainly on its time evolution, the consistency between the semi-analytical treatment and the numerical simulation ensures that the result we obtained from semi-analytical approach is reasonable.

\begin{figure}[h]
\includegraphics[width=0.45\textwidth]{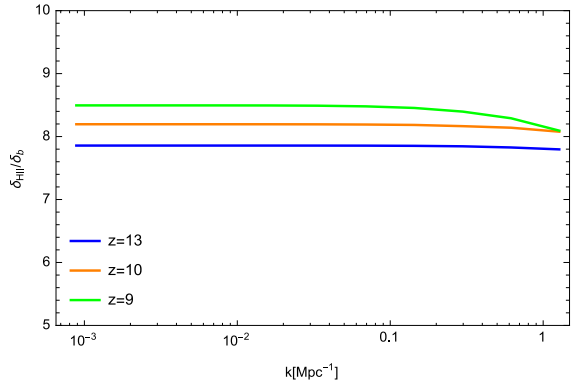}
\caption{\label{bias} The bias between $\delta_{HII}$ and $\delta_b$ in the standard $\Lambda$CDM model obtained from the semi-analytical model.}
\end{figure}

However at small scales (bigger $k$), the difference between the semi-analytical treatment and the numerical simulation appears. This is because that in semi-analytical models the bias of $\delta_{HII}$ and $\delta_b$ reduce slowly with the increase of $k$, as shown in Fig. \ref{bias}, whereas they drop quickly in simulations (see Fig. 6 in \cite{Zahn2011}). The deep reason of such difference could be attributed to the fact that the semi-analytic model based on linear theory may break down in small scales. 

In addition, there are many parameters which can influence physics in this epoch, such as the formation of the first generation of stars or galaxies and their spatial distribution, the energy spectra of the ionizing photons of different types of ionizing sources (stars, quasars), dynamics of gas under the influence of radiation, etc.. Whether these parameters can alter the ionization history, and alleviate the difference between semi-analytic approach and numerical simulation is a question we want to understand in the future. But a clear answer of these parameters on the role of physics in the reionization epoch can only be got after we have measurements of the high-redshift 21cm spectra.\\

\end{appendices}

\bibliography{ide21}

\end{document}